\documentclass[12pt]{article}
\usepackage{cite}
\usepackage{stackrel}
\usepackage{hyperref}
\usepackage{graphicx}
\graphicspath{{./}}

\newcommand{\nc}{N_c}
\newcommand{\nf}{n_f}
\newcommand{\nl}{n_l}
\newcommand{\nh}{n_h}
\newcommand{\si}{\mbox{si}}
\newcommand{\cf}{C_F}
\newcommand{\ca}{C_A}
\newcommand{\dR}{d_R}
\newcommand{\dabc}{d^{abc}}
\newcommand{\tr}{T_F}
\newcommand{\as}{a_s}
\newcommand{\asT}{\breve{a}_{s}}
\newcommand{\asH}{\tilde{a}_{s}}
\newcommand{\z}[1]{\zeta_{#1}}
\newcommand{\A}[1]{a_{#1}}
\newcommand{\Log}[1]{\ln\left(#1\right)}
\newcommand{\LmusdmMSbars}{\ell_{\mu}}
\newcommand{\logtwo}{l_{2}}
\newcommand{\MSbar}{\overline{\mbox{MS}}}
\newcommand{\vep}{\varepsilon}

\begin{document}    

\begin{titlepage}
\noindent
\hfill Date: April 13, 2014
\mbox{}

\vspace{0.5cm}
\begin{center}
  \begin{Large}
    \begin{bf}
      Higher order QCD results for the fermionic
      contributions of the Higgs-boson decay into two
      photons and the decoupling function for the $\MSbar$
      renormalized fine-structure constant 
    \end{bf}
  \end{Large}
  \vspace{0.8cm}

  \begin{large}
    Christian Sturm
\footnote{\href{Christian.Sturm@physik.uni-wuerzburg.de}{Christian.Sturm@physik.uni-wuerzburg.de}}
  \end{large}
  \vskip .7cm
	{\small {\em 
Universit{\"a}t W{\"u}rzburg,\\
Institut f{\"u}r Theoretische Physik und Astrophysik,\\
Emil-Hilb-Weg 22,\\
D-97074 W{\"u}rzburg, \\
Germany}}\\
	\vspace{0.8cm}
\vspace*{7cm}
{\bf Abstract}
\end{center}
\begin{quotation}
  \noindent
  We compute the decoupling function of the $\MSbar$ renormalized
  fine-structure constant up to four-loop order in perturbative QCD. The
  results are used in order to determine the related top-quark
  contributions to the Higgs-boson decay into two photons in the heavy
  top-quark mass limit to order $\alpha_s^4$.
\end{quotation}
\end{titlepage}

\section{Introduction\label{sec:Introduction}}
The discovery of a Higgs-particle~\cite{Chatrchyan:2012ufa,Aad:2012tfa}
which is, within the current uncertainties, consistent with the
expectations of the Standard Model (SM) Higgs boson ($H$) was a big
success of the Large Hadron Collider (LHC) experiments.  It is now
interesting and important to further understand and study its properties
also from theory side.

The dominant production mechanism for the SM Higgs boson at the LHC is
the gluon fusion process, whereas the Higgs-boson decay into two
photons, $H\to\gamma\gamma$, provides a clean channel for the study of
its properties, not only at the LHC, but also at a future linear
collider. As a result of this these processes are studied at high order
in perturbation theory to obtain most accurate predictions. From theory
point of view both reactions are loop induced processes, since the SM
Higgs boson couples only to massive particles, so that the leading order
processes are here already at the one-loop level.  In the limit in which
the heavy top-quark mass $m_t$ is much larger than the mass of the Higgs
boson, $m_H\ll 2\*m_t$, the fermionic contributions can be described by
an effective coupling using a low-energy theorem
(LET)~\cite{Ellis:1975ap, Shifman:1979eb, Vainshtein:1980ea,
  Kniehl:1995tn}.

Within this work we study higher order corrections in quantum
chromodynamics (QCD) to the decay $H\to\gamma\gamma$. For electroweak
corrections we refer to
Refs.~\cite{Aglietti:2004nj,Fugel:2004ug,Degrassi:2005mc,Passarino:2007fp,Actis:2008ts}. The
two-loop QCD corrections are known since long and have been determined
first in the heavy top-quark mass limit in
Refs.~\cite{Zheng:1990qa,Dawson:1992cy}. The full top-quark mass
dependence has then been obtained numerically and analytically in
Refs.~\cite{Djouadi:1990aj, Melnikov:1993tj, Inoue:1994jq, Spira:1995rr,
  Fleischer:2004vb,Harlander:2005rq,Aglietti:2006tp}.
In the following we will focus on the purely fermionic contributions to
the decay $H\to\gamma\gamma$ which form a gauge-invariant set of
diagrams. We will further distinguish between the singlet and
non-singlet contributions. The latter are given by those diagrams for
which the Higgs boson and the photons couple to the same top-quark loop.
The three-loop QCD corrections of the non-singlet diagrams including
power corrections up to $\mathcal{O}((m_H^2/(4\*m_t^2))^2)$ were
determined in Ref.~\cite{Steinhauser:1996wy}. This result was
complemented in Ref.~\cite{Maierhofer:2012vv}, where additional power
corrections of higher order in $m_H^2/(4\*m_t^2)$ as well as the
complete singlet contribution has been computed.  The symbol $m_H$ is
here the mass of the Higgs boson and $m_t$ is the mass of the top quark.

The four-loop QCD corrections induced by a heavy quark were derived in
the heavy quark mass limit by using the known results for the
$\beta$-function and mass anomalous dimensions in
Ref.~\cite{Chetyrkin:1997un}.  In the following we will first compute
the QCD corrections to the decoupling function of the QED coupling
constant up to four-loop order.  This extends the result of
Ref.~\cite{Chetyrkin:1997un} by one order in perturbation theory. As an
application we will use this result in order to derive independently the
four-loop QCD corrections induced by a heavy quark to the decay
$H\to\gamma\gamma$ in the heavy top-quark mass limit by applying the
LET, instead of using the anomalous dimensions as in
Ref.~\cite{Chetyrkin:1997un}. In the next step we will in turn exploit
the renormalization group equation (RGE) in order to reconstruct the
logarithmic part of the vacuum polarization function at five-loop order,
which is again sufficient in order to derive the corresponding
contributions to the decay amplitude $H\to\gamma\gamma$ to order
$\alpha_s^{4}$; here $\alpha_s$ is the strong coupling constant.

This paper is structured as follows: In the next Section~\ref{sec:gen}
we discuss the generalities and notations which are needed for this
work.  In Section~\ref{sec:results} we describe the calculation and
present our results. Finally we close with a short summary in
Section~\ref{sec:DiscussConclude}. In the Appendix we
provide supplementary information.

\section{Generalities and notation\label{sec:gen}}
\begin{figure}[ht!]
\begin{center}
\begin{minipage}{3cm}
\begin{center}
\includegraphics[bb=72 354 540 720,width=3cm]{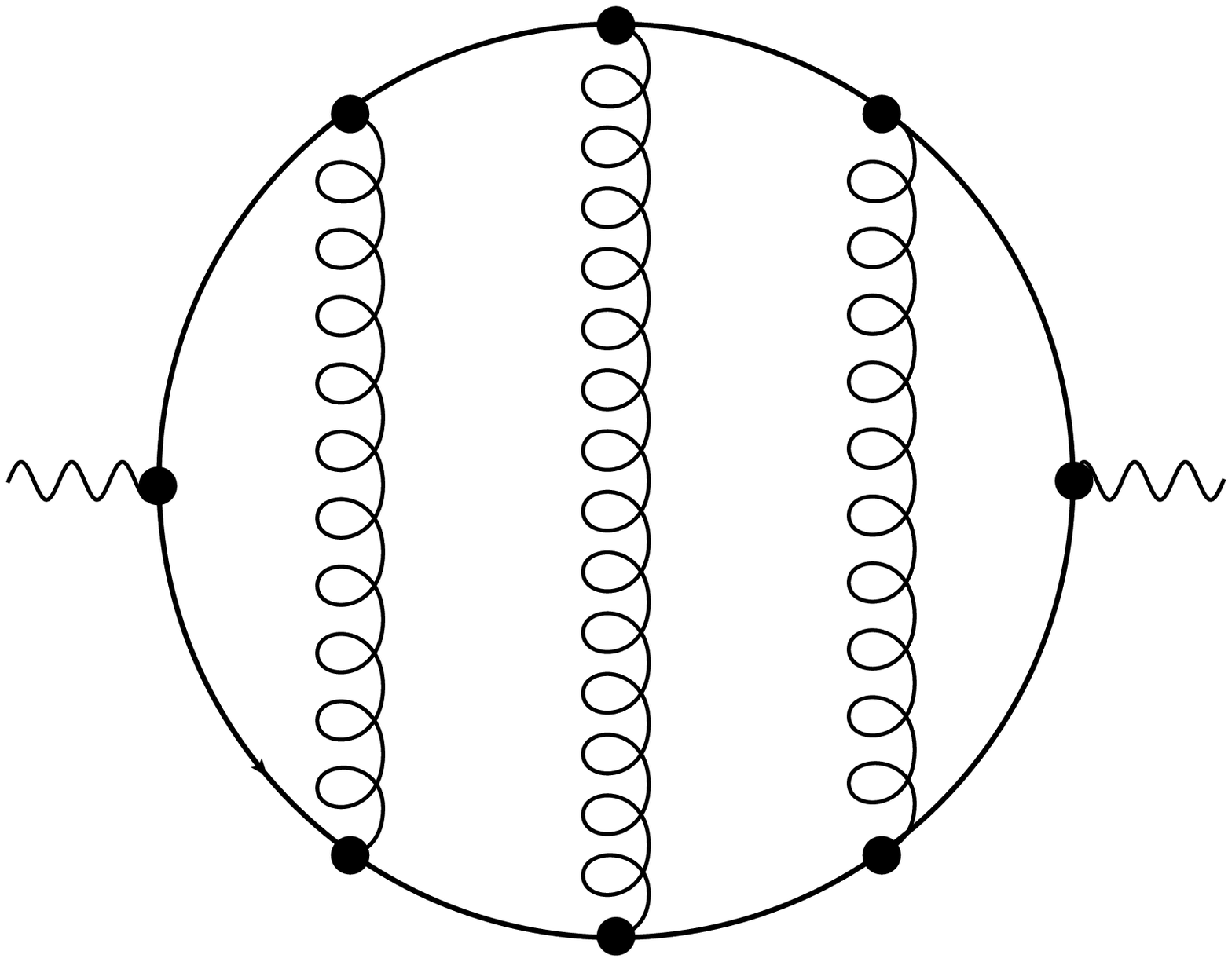}\\
$Q_t^2\cf^3$
\end{center}
\end{minipage}
\begin{minipage}{3cm}
\begin{center}
\includegraphics[bb=72 366 540 720,width=3cm]{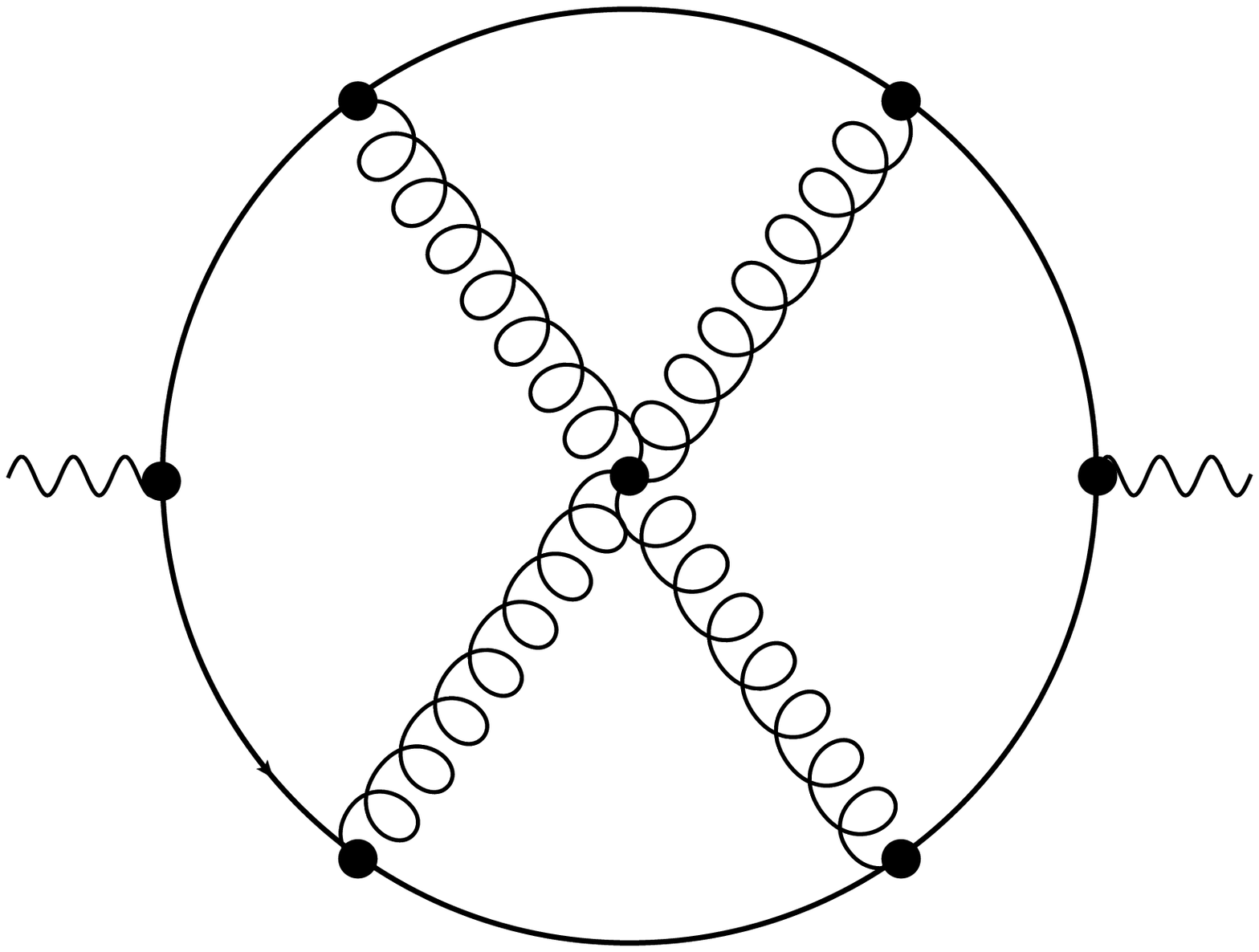}\\
$Q_t^2\cf^2\ca$
\end{center}
\end{minipage}
\begin{minipage}{3cm}
\begin{center}
\includegraphics[bb=72 354 540 720,width=3cm]{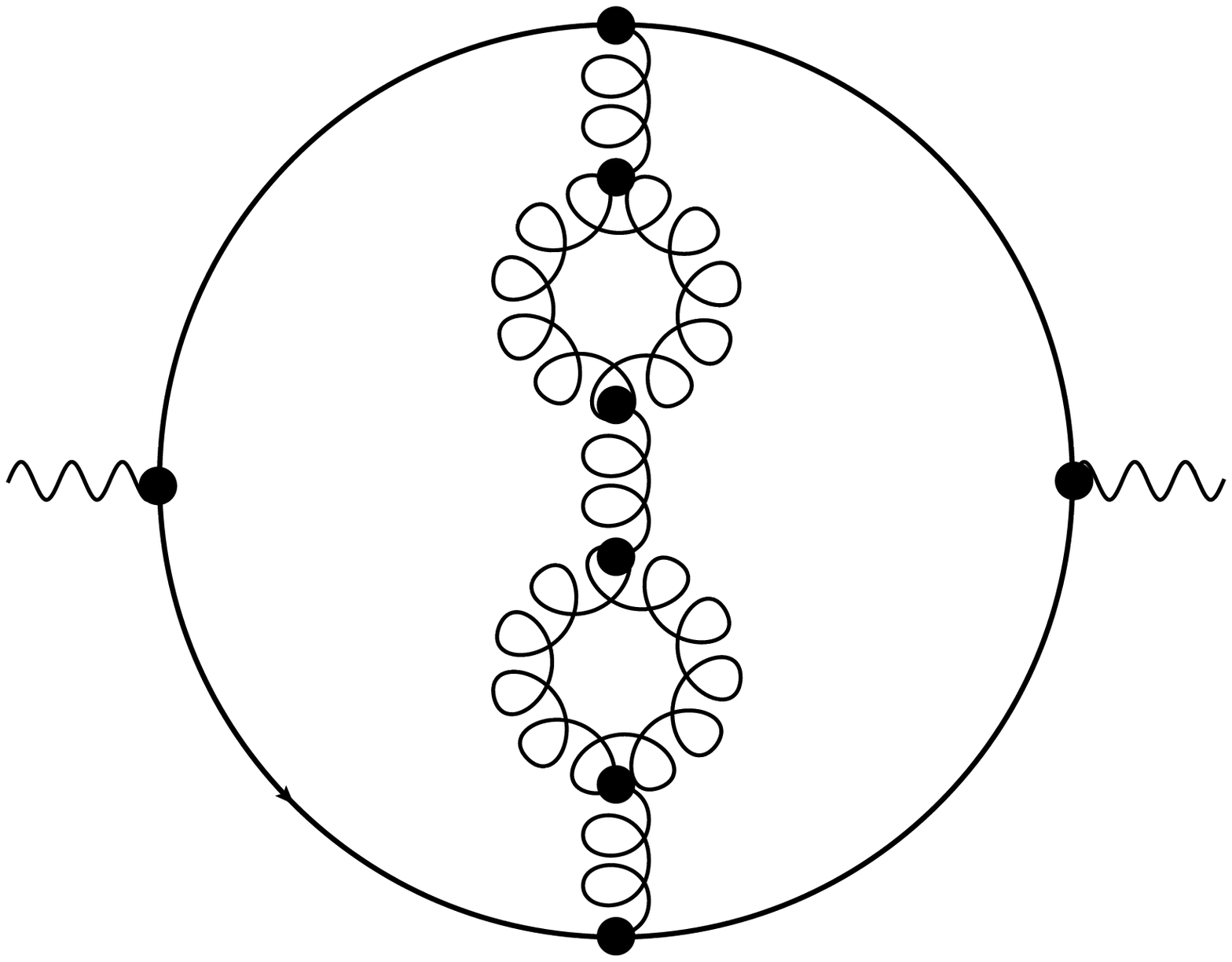}\\
$Q_t^2\cf\ca^2$
\end{center}
\end{minipage}
\begin{minipage}{3cm}
\begin{center}
\includegraphics[bb=72 366 540 720,width=3cm]{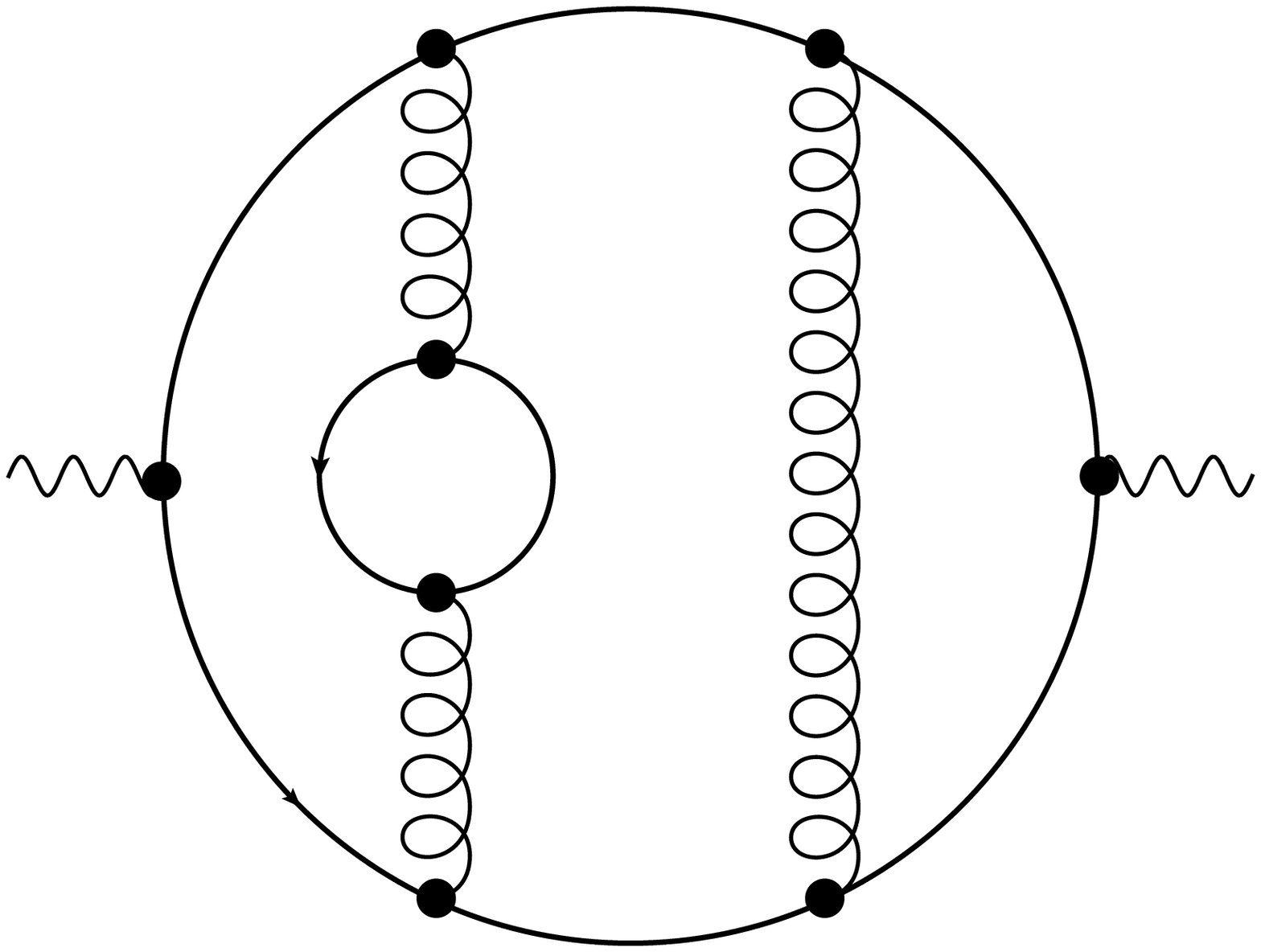}\\
$Q_t^2\nh\cf^2\tr$
\end{center}
\end{minipage}\\[2ex]
\begin{minipage}{3cm}
\begin{center}
\includegraphics[bb=72 354 540 720,width=3cm]{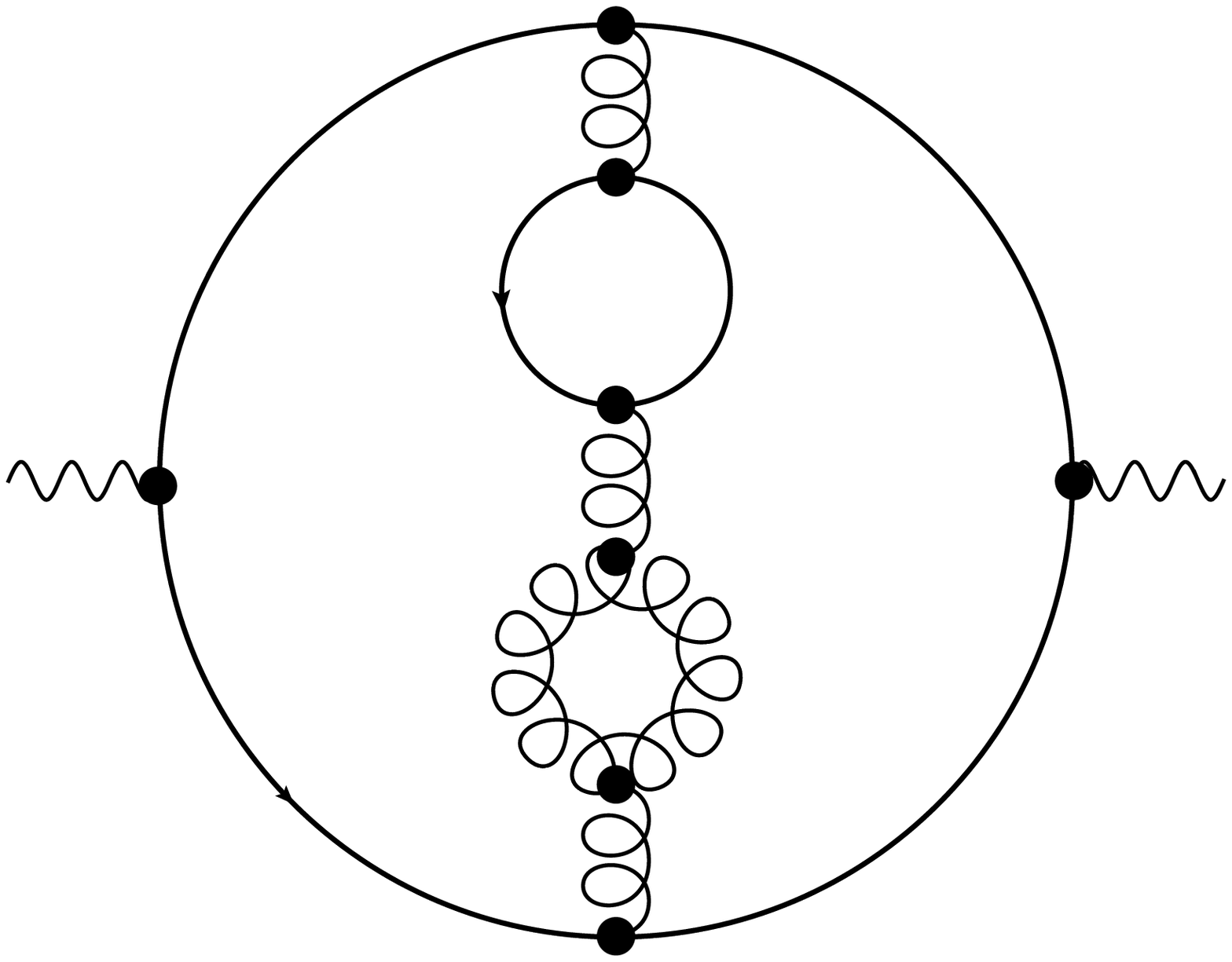}\\
$Q_t^2\nh\cf\ca\tr$
\end{center}
\end{minipage}
\begin{minipage}{3cm}
\begin{center}
\includegraphics[bb=72 366 540 720,width=3cm]{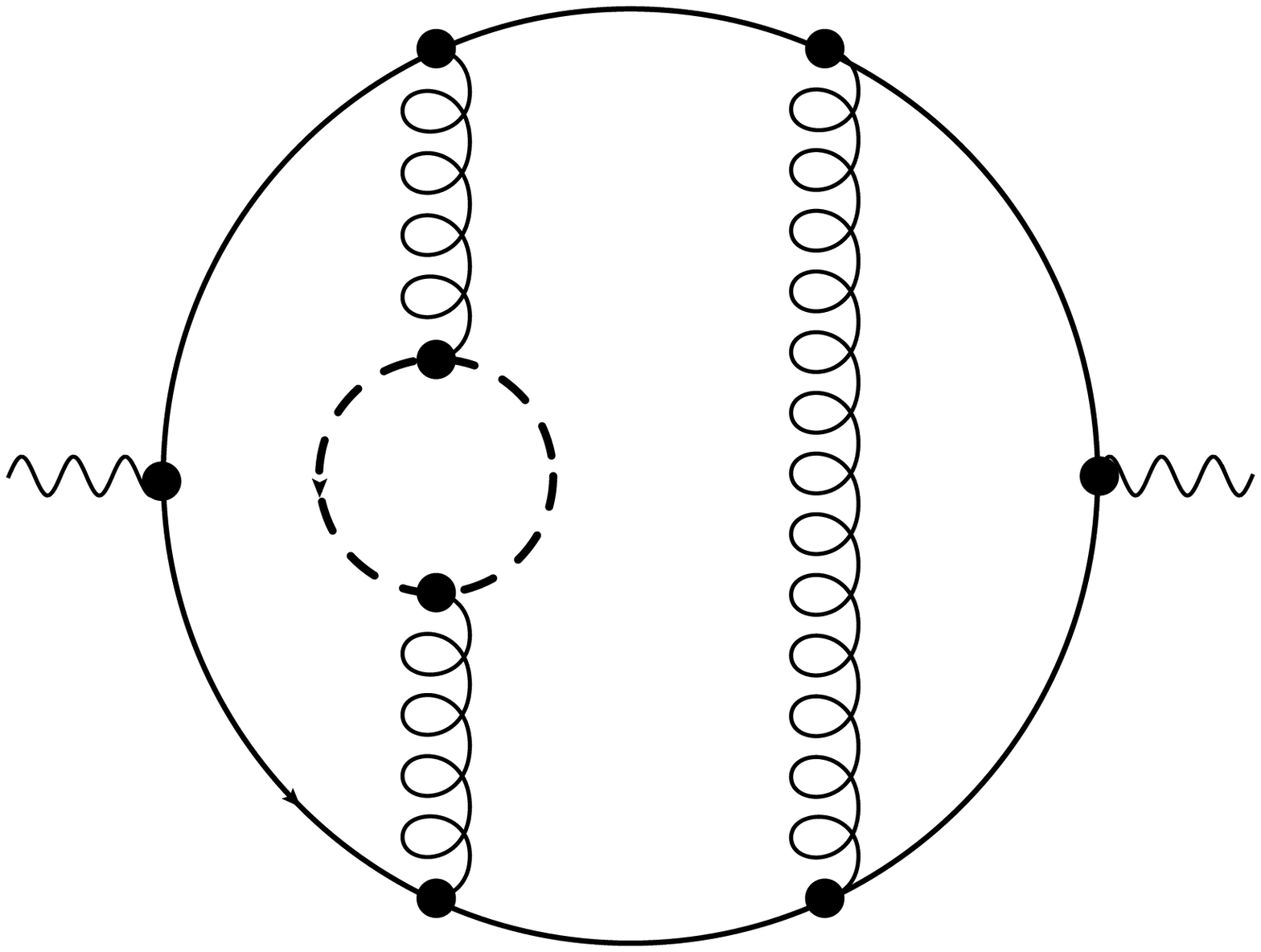}\\
$Q_t^2\nl\cf^2\tr$
\end{center}
\end{minipage}
\begin{minipage}{3cm}
\begin{center}
\includegraphics[bb=72 354 540 720,width=3cm]{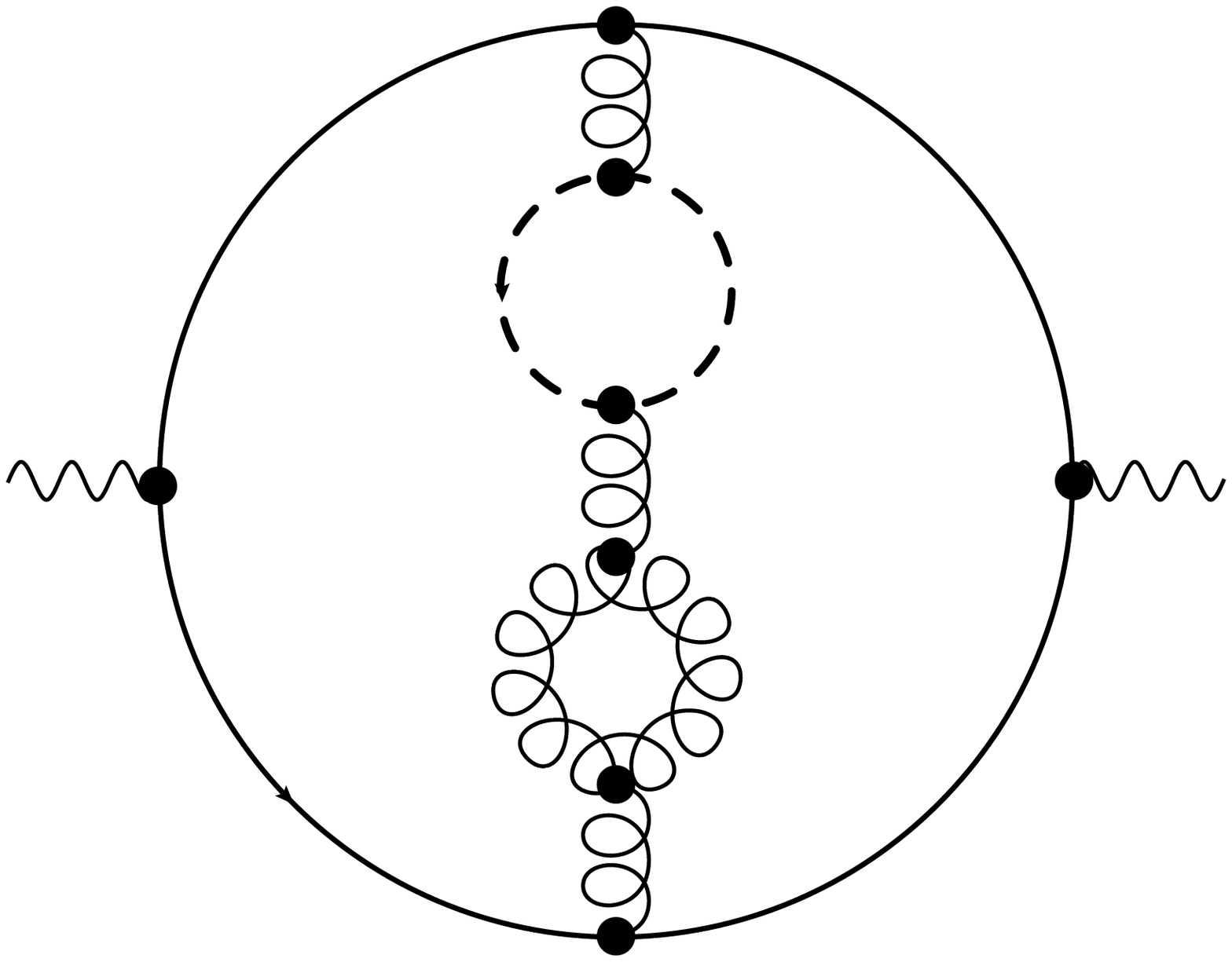}\\
$Q_t^2\nl\cf\ca\tr$
\end{center}
\end{minipage}
\begin{minipage}{3cm}
\begin{center}
\includegraphics[bb=72 354 540 720,width=3cm]{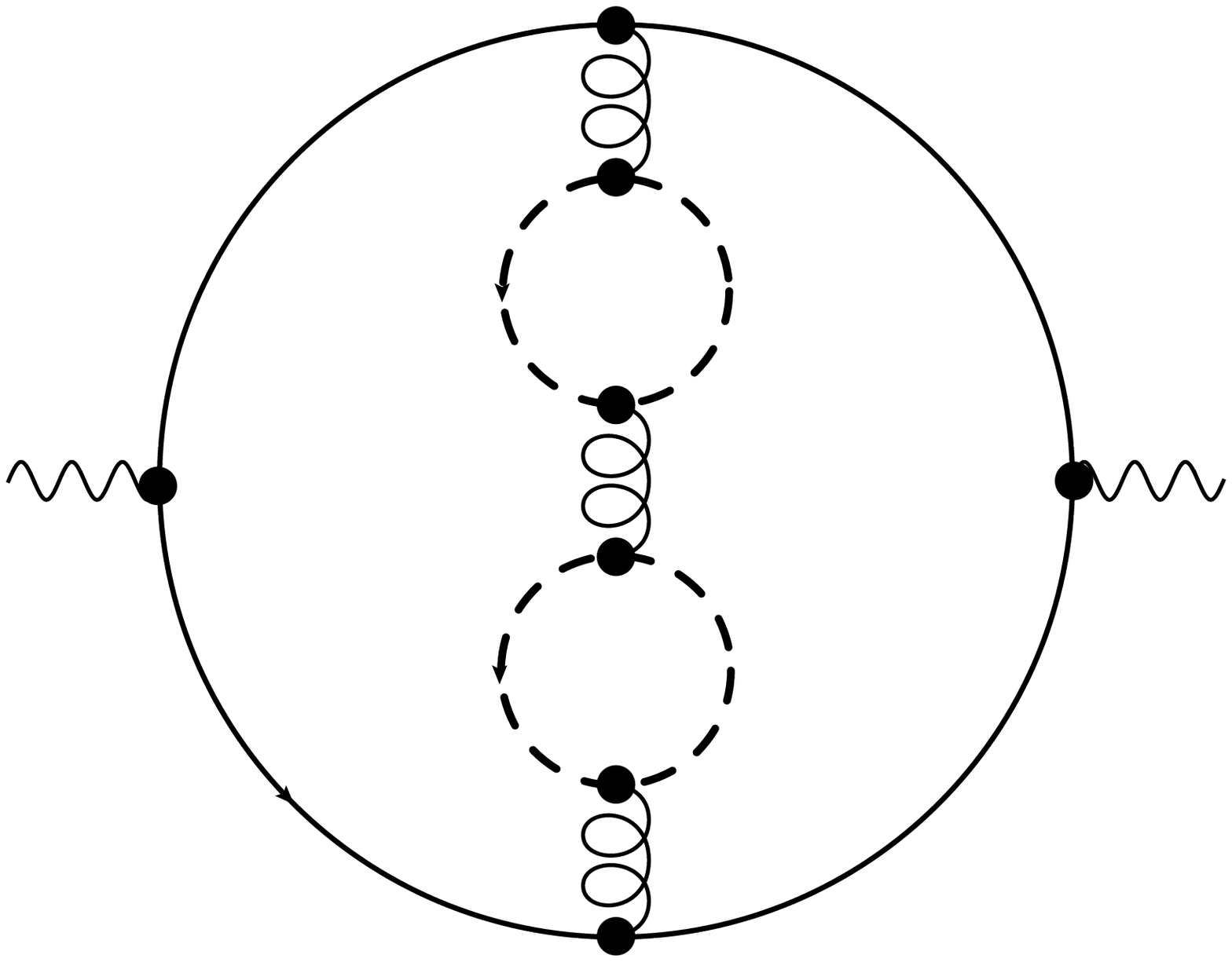}\\
$Q_t^2\nl^2\cf\tr^2$
\end{center}
\end{minipage}\\[2ex]
\begin{minipage}{3cm}
\begin{center}
\includegraphics[bb=72 354 540 720,width=3cm]{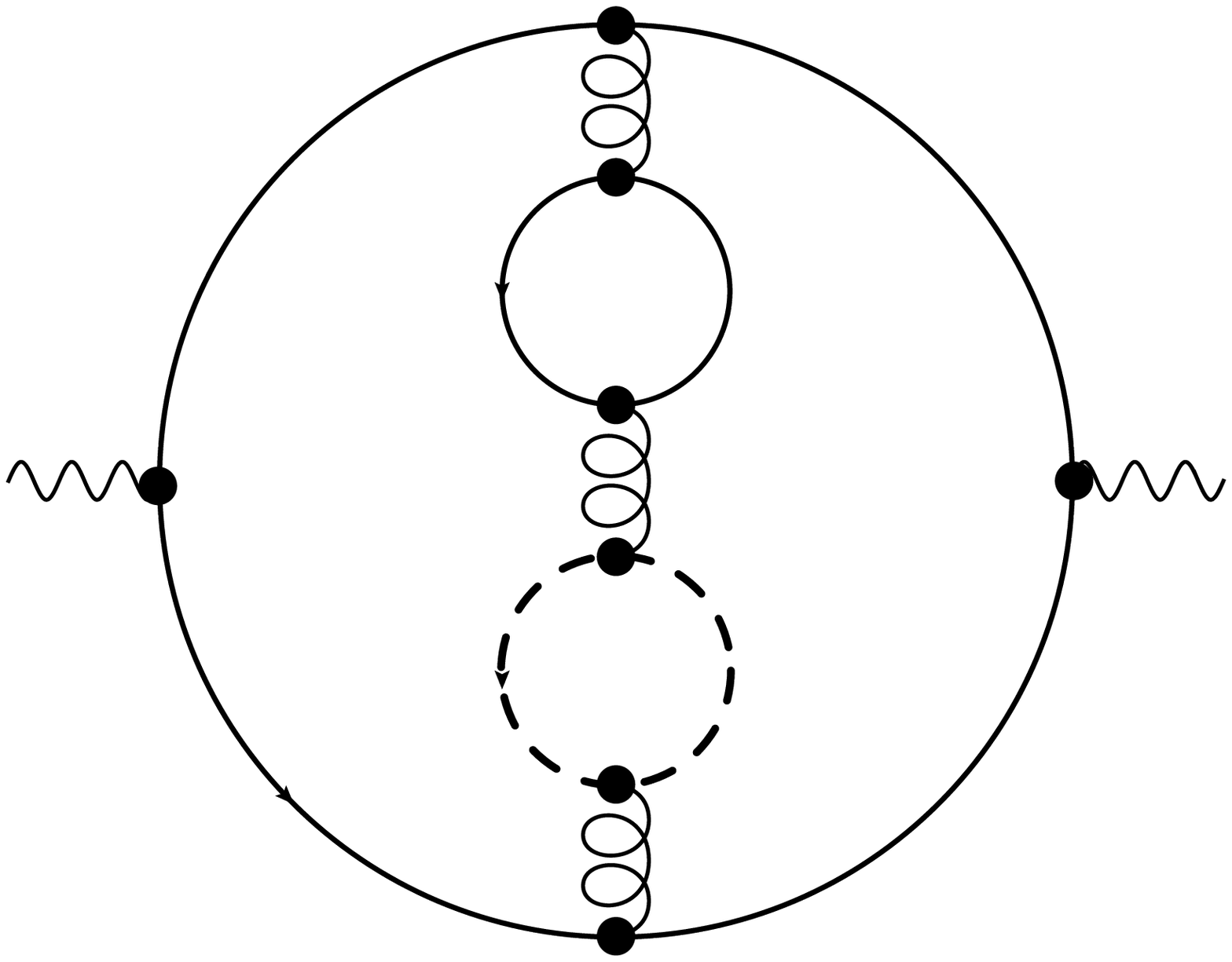}\\
$Q_t^2\nh\nl\cf\tr^2$
\end{center}
\end{minipage}
\begin{minipage}{3cm}
\begin{center}
\includegraphics[bb=72 354 540 720,width=3cm]{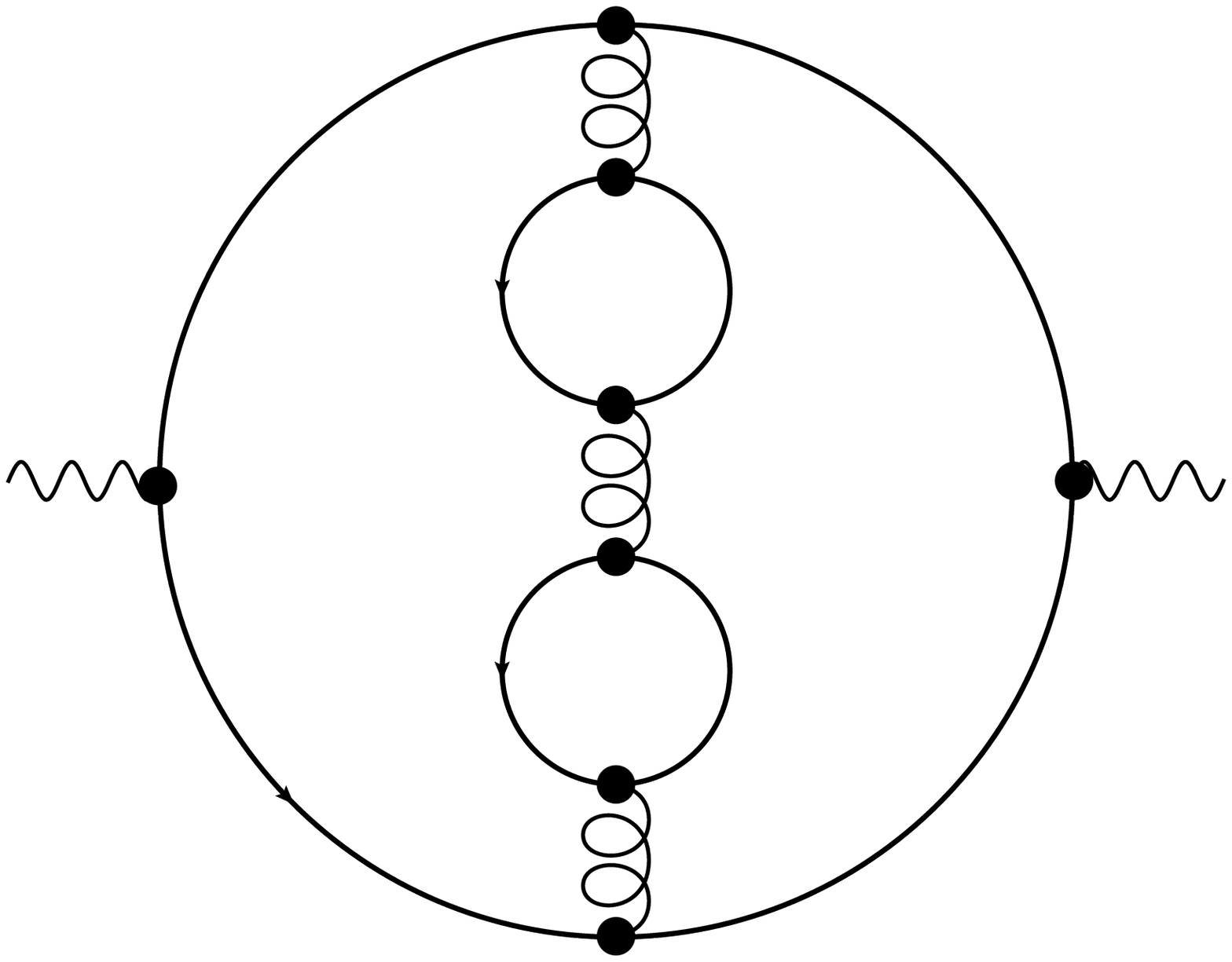}\\
$Q_t^2\nh^2\cf\tr^2$
\end{center}
\end{minipage}
%
%
\begin{minipage}{3cm}
\begin{center}
\includegraphics[bb=72 352 540 720,width=3cm]{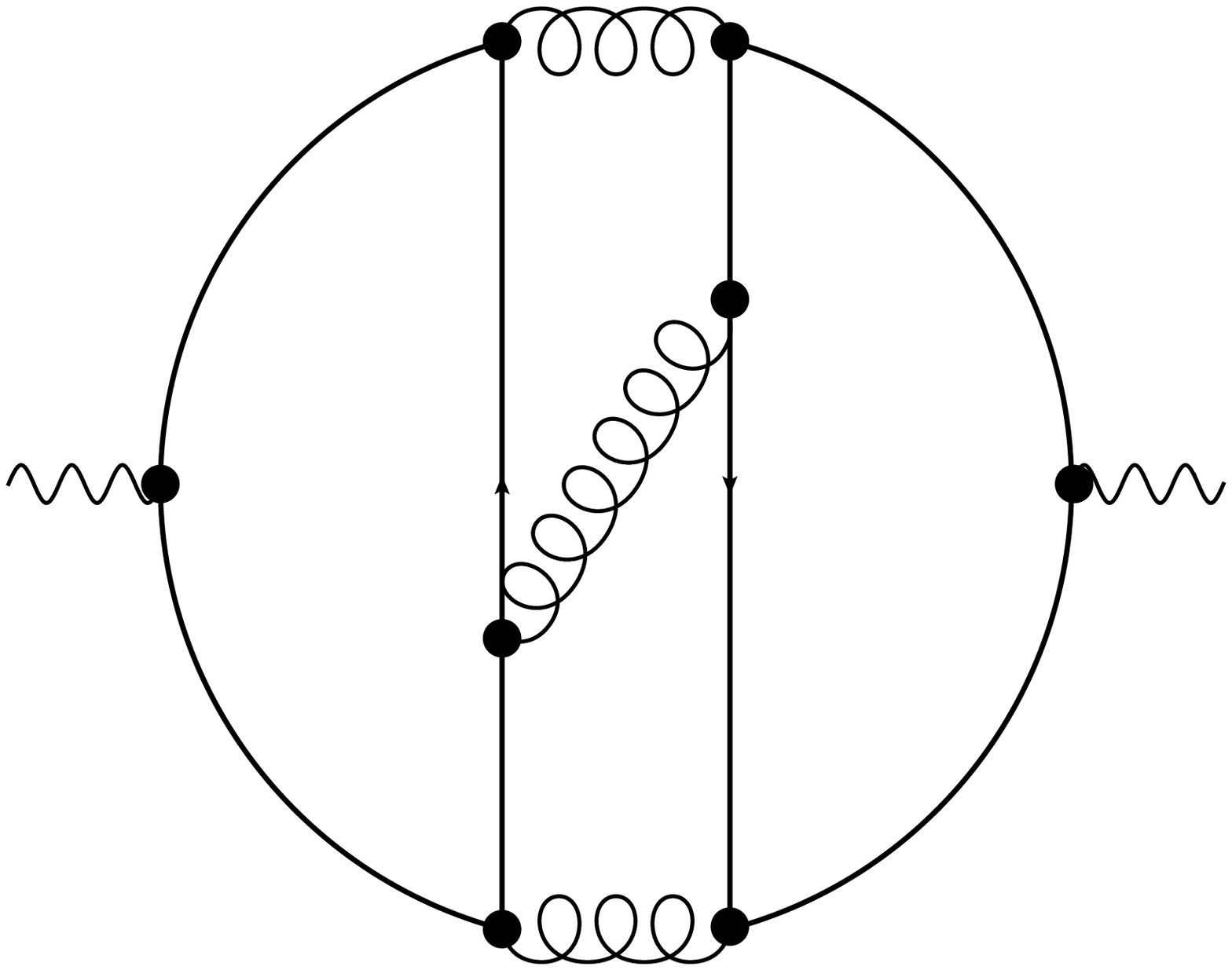}\\
$Q_t^2\si\,\dabc\dabc$
\end{center}
\end{minipage}
\begin{minipage}{3cm}
\begin{center}
\includegraphics[bb=72 366 540 720,width=3cm]{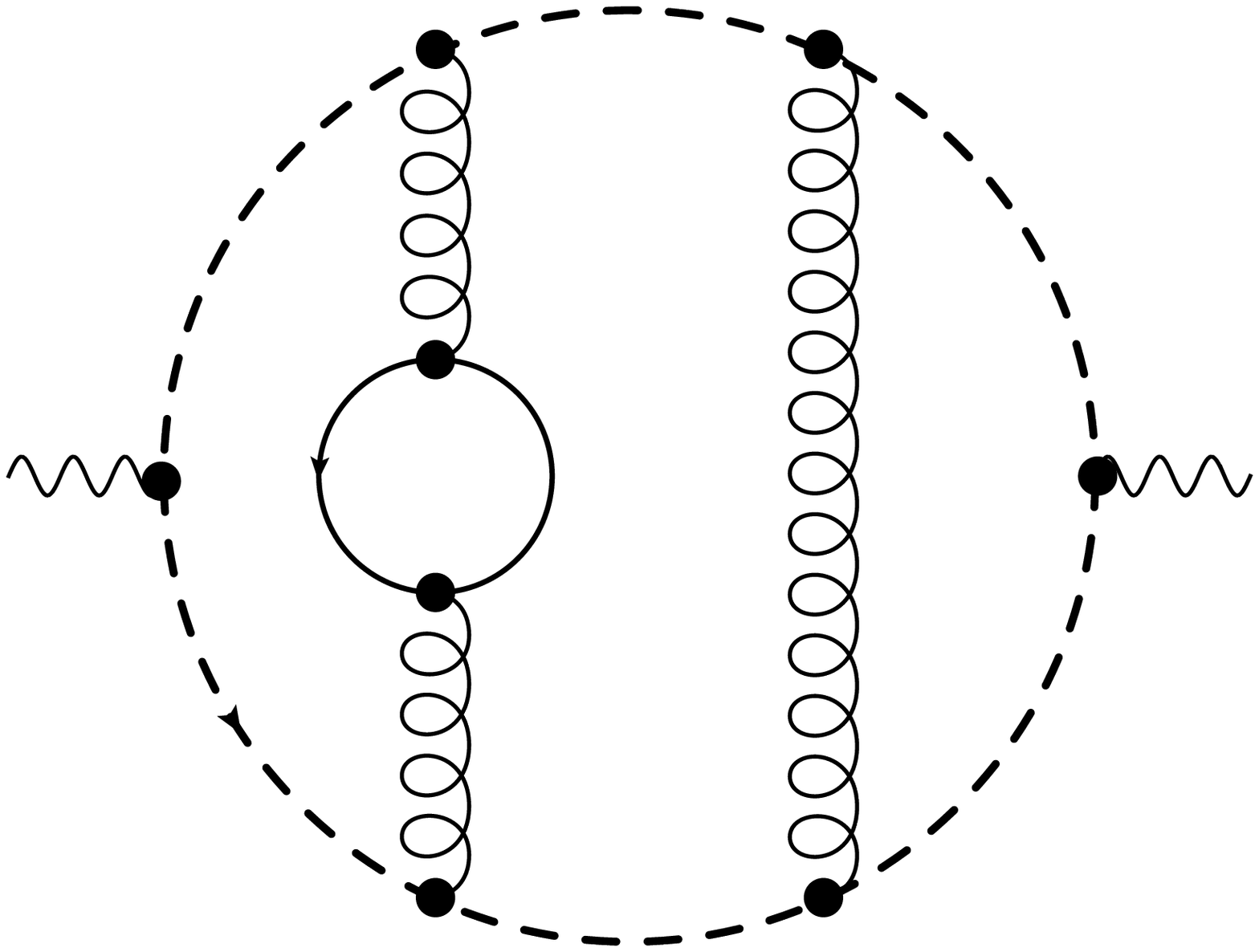}\\
$Q_q^2\,\nh\cf^2\tr$
\end{center}
\end{minipage}\\[2ex]
\begin{minipage}{3cm}
\begin{center}
\includegraphics[bb=72 354 540 720,width=3cm]{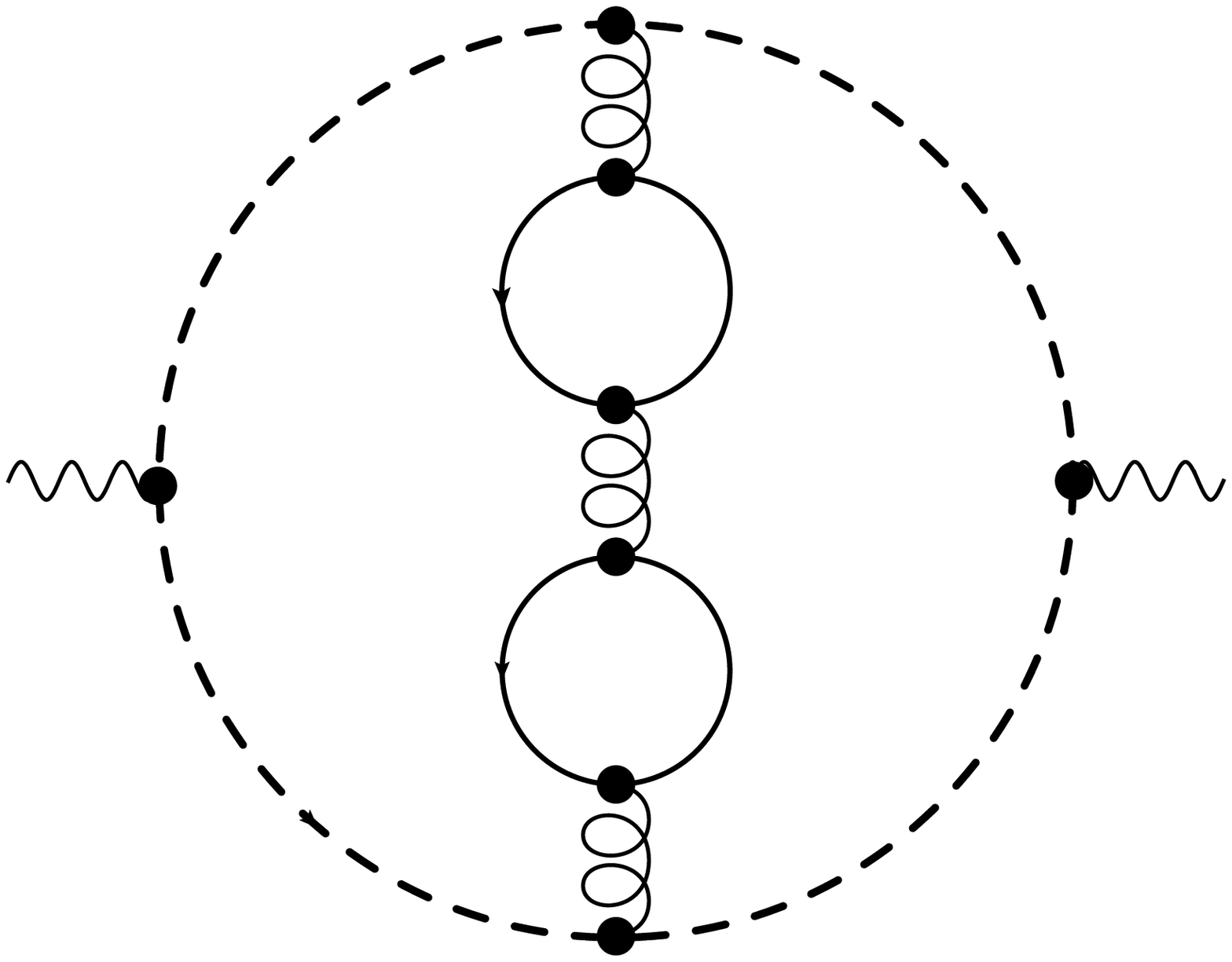}\\
$Q_q^2\,\nh^2\cf\tr^2$
\end{center}
\end{minipage}
\begin{minipage}{3cm}
\begin{center}
\includegraphics[bb=72 354 540 720,width=3cm]{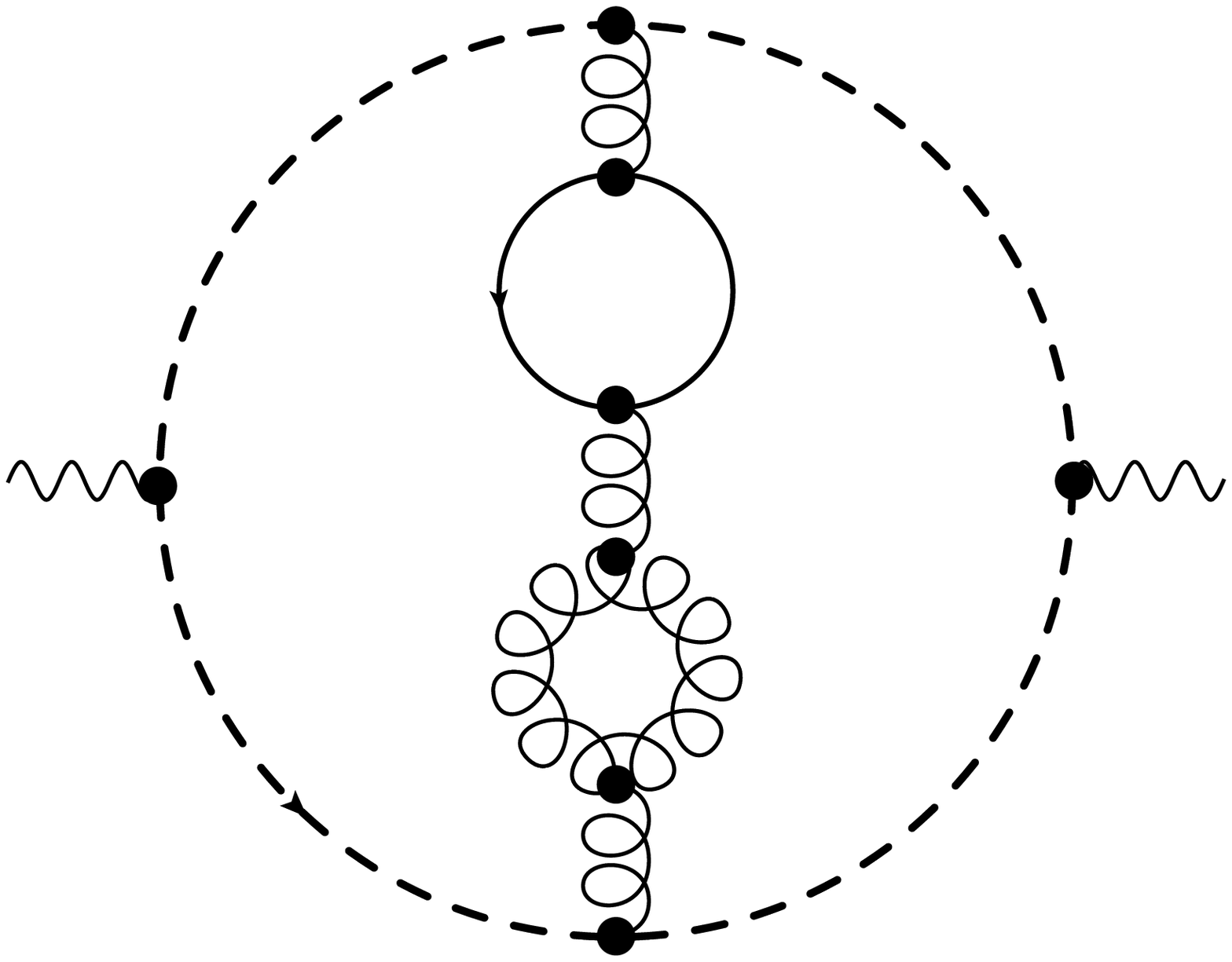}\\
$Q_q^2\,\nh\cf\ca\tr$
\end{center}
\end{minipage}
\begin{minipage}{3cm}
\begin{center}
\includegraphics[bb=72 354 540 720,width=3cm]{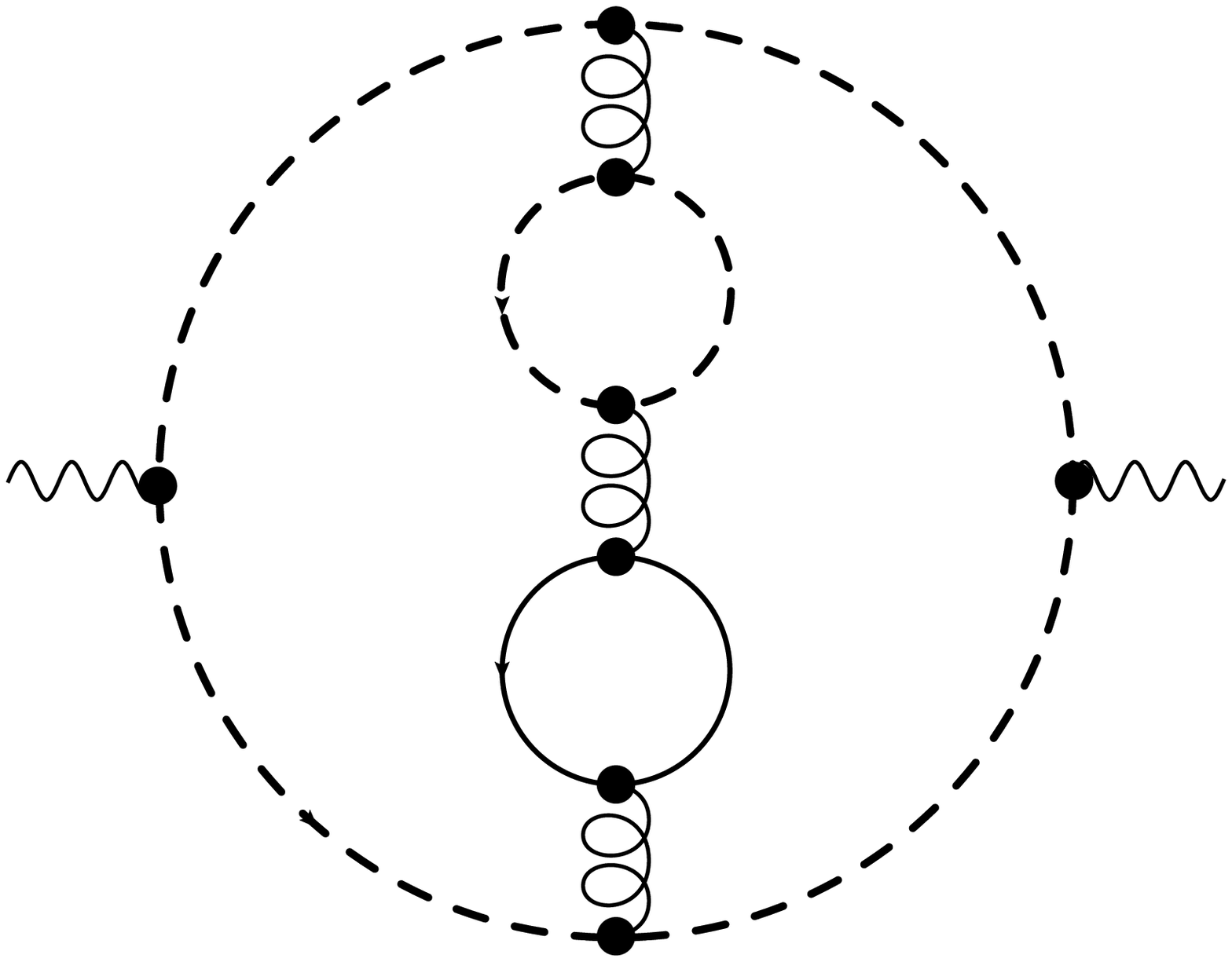}\\
$Q_q^2\,\nh\nl\cf\tr^2$
\end{center}
\end{minipage}
\begin{minipage}{3cm}
\begin{center}
\includegraphics[bb=72 352 540 720,width=3cm]{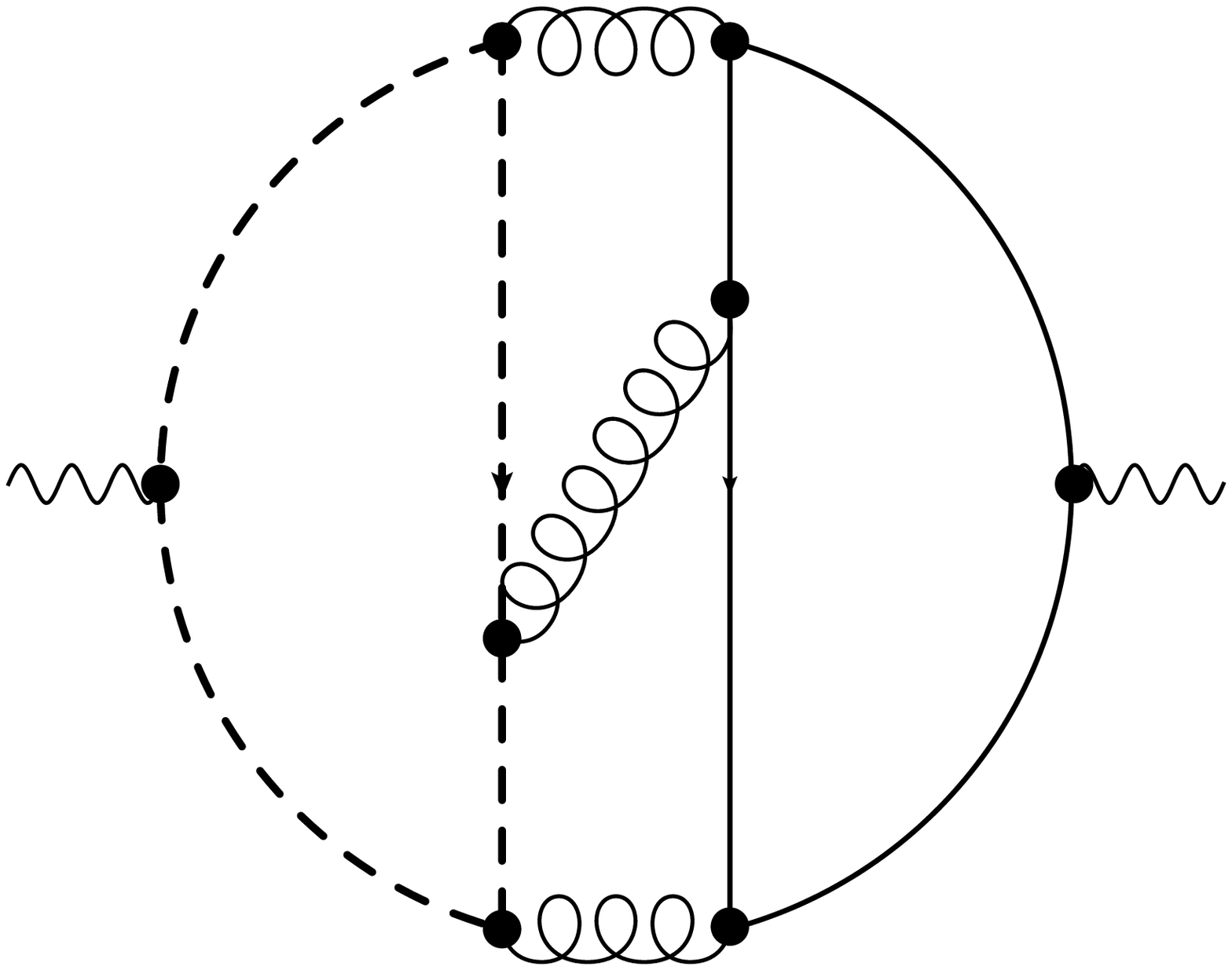}\\
$Q_q\,Q_t\,\si\,\dabc\dabc$
\end{center}
\end{minipage}
\end{center}
\caption{For each color structure which appears at four-loop order in
  the vacuum polarization function $\Pi^{0h}(q^2=0,m_t^0)$ one example
  Feynman graph is shown. The twisted lines denote gluons, the solid
  lines are heavy top-quarks and the dashed lines represent massless
  quarks. The number of inserted heavy and light fermion loops as well
  as the charge factors are given in the subscripts. The color factors
  are defined in Section~\ref{sec:results}.\label{fig:Pi4loop}}
\end{figure}

The partial decay width for the Higgs-boson decay into two photons is at
leading order given by
\begin{equation}
\label{eq:decaywidth}
\Gamma(H\to\gamma\gamma)={m_H^3\over64\*\pi}\*\Big|A_W(\tau_W)+\sum_{f}A_f(\tau_f)\Big|^2\,,
\end{equation}
where $A_W(\tau_W)$ is the contribution which arises from purely bosonic
diagrams and $A_f(\tau_f)$ is the fermionic contribution to the
amplitude, respectively. The symbol $\tau_i$ denotes the mass ratio
$m_H^2/(4\*m_i^2)$ ($i=W,f$), where $m_W$ is the mass of the $W$-boson
and $m_f$ is the mass of a heavy fermion.  Within this work we focus
only on the term $A_f(\tau_f)$, since we will consider QCD at higher
order in perturbation theory.  In particular the fermionic contribution
of the amplitude is dominated by the contribution which originates from
the top quark, $A_t(\tau_t)$ ($f=t$), since the top quark is the
heaviest fermion of the SM. In the limit of a heavy top-quark mass,
$m_t\to \infty$, the leading order result for the amplitude reads
$\hat{A}_t=\nc\*{2\*Q_t^2\*\alpha\over3\*\pi\*v}$, with
$v=2^{-1/4}G_F^{-1/2}$. The symbol $\nc$ is here the number of colors of
$SU(\nc)$, $Q_t$ is the electric charge factor of the top quark,
$\alpha$ is the fine-structure constant, and $G_F$ is the Fermi-coupling
constant.  By integrating out the heavy top-quark one can construct a
heavy top-quark effective Lagrangian which describes the interactions of
the Higgs field $H$ with the photon field and the $\nl$ light quark
flavors, which are considered as massless
\begin{equation}
\label{eq:Lhgg}
\mathcal{L}_{\mbox{\scriptsize{eff}}}=-{H^0\over v^0}F^{'0,\mu\nu}F^{'0}_{\mu\nu}\,C^0_{1\gamma}\,.
\end{equation}
The symbol $v^0$ is the vacuum expectation value and $F^{'0}_{\mu\nu}$
is the field strength tensor.  The subscript $0$ denotes here and in the
following a bare quantity and the prime implies quantities in the
effective theory with $\nl$ light active quark flavors.  The coefficient
function
$C^0_{1\gamma}=-{1\over2}\*m_{t}^{0}{\partial\ln(\zeta_{g\gamma}^0)\over\partial
  m_{t}^0}$ depends on the decoupling function $\zeta^0_{g\gamma}$,
which relates, after renormalization, the $\MSbar$ renormalized
fine-structure constant in the effective and full theory with
$\nf=\nl+1$ active quark flavors:
$\overline{\alpha}'(\mu)=\zeta^2_{g\gamma}\overline{\alpha}^{(\nf)}(\mu)$.
The decoupling function $\zeta^0_{g\gamma}$ can be determined by the
computation of the hard part of the photon vacuum polarization function
$\Pi^{0h}(q^2,m_t^0)$ at zero momentum, $q^2=0$, e.~g. by considering
only those diagrams which involve the heavy top quark,
\begin{equation}
{\zeta^0_{g\gamma}}^2={1\over 1+\Pi^{0h}(q^2=0,m_t^0)}\,.
\end{equation}
Fig.~\ref{fig:Pi4loop} shows some example Feynman diagrams which
contribute to the calculation of $\Pi^{0h}(q^2=0,m_t^0)$.
In the following we decompose the hard part of the vacuum polarization
function into three contributions:
\begin{equation}
\Pi^{0h}(q^2=0,m_t^0)=\Pi_{t}^{0h}(q^2=0,m_t^0)+\sum_{i=1}^{\nl}\Pi_{q_{i}}^{0h}(q^2=0,m_t^0)+\sum_{i=1}^{\nl}\Pi_{tq_{i}}^{0h}(q^2=0,m_t^0)\,. 
\end{equation}
The first term $\Pi_{t}^{0h}(q^2=0,m_t^0)$ is the part which is
proportional to the charge factor $Q_t^2$ and arises from the diagrams
in which the external photons couple only to top quarks. In contrast,
the second term $\sum_{i=1}^{\nl}\Pi_{q_i}^{0h}(q^2=0,m_t^0)$ arises
starting from three-loop order in QCD from those contributions for which
the external photons couple to light quarks with the global charge
factor $Q_{q_i}^2$.  The last term arises from singlet diagrams which
are characterized by the fact that one photon couples to a light quark
loop and the other photon to a massive top-quark loop. This last term is
proportional to the charge factors $Q_t\*Q_{q_i}$ and appears for the
first time at four-loop order.

For the decay process $H\to\gamma\gamma$ we will in the following
consider only those gauge-invariant contributions for which the photons
couple directly to a top-quark loop.  Making Eq.~(\ref{eq:Lhgg})
explicit one obtains for the effective Higgs-photon-photon coupling the
Lagrangian
\begin{equation}
\label{eq:LHgamgam}
\mathcal{L}_{H\gamma\gamma}=-{1\over4}F^{0,\mu\nu}F^{0}_{\mu\nu}\,
                           m^{0}_t{\partial\over\partial
                             m^{0}_t}\Pi_{t}^{0h}(0,m_t^0){H^0\over v^0}\,.
\end{equation}
As a result of this one can determine from the hard part of the photon
vacuum polarization function the leading top-quark contribution of the
bare amplitude $A^{0,\infty}_t=-{m^0_t\over
  v}{\partial\over\partial{m^0_t}}{\Pi}_{t}^{0h}(0,m^0_t)$. Considering
QCD corrections the $\MSbar$ renormalized vacuum polarization function
obeys the RGE
\begin{equation}
\label{eq:anodim}
\left[
\beta(a_s)\*{\partial\over\partial a_s}+
\gamma_m\*\overline{m}_t\*{\partial\over\partial \overline{m}_t}+
     \mu^2\*{\partial\over\partial\mu^2}
\right]\*
\overline{\Pi}_{t}^{0h}(q^2=0,\overline{m}_t,a_s,\mu)=
4\*\pi\*\alpha\*Q_t^2\*\gamma_{VV}\,,
\end{equation}
with $\as={\alpha_s^{(\nf)}(\mu)/\pi}$, where $\alpha_s^{(\nf)}(\mu)$
and $\overline{m}_t\equiv\overline{m}_t(\mu)$ are the strong coupling
constant and the top-quark mass renormalized in the $\MSbar$
scheme. Both depend on the renormalization scale
$\mu$. Eq.~(\ref{eq:anodim}) can be used in order to check the scale
dependent part of
$\overline{\Pi}_{t}^{0h}(q^2=0,\overline{m}_t,a_s,\mu)$.  The anomalous
dimension $\gamma_{VV}$ has been computed in Ref.~\cite{Baikov:2012zm}
up to five-loop order.  The mass anomalous dimension $\gamma_m={\partial
  \ln\overline{m}_t\over\partial\ln\mu^2}$ as well as the QCD
$\beta$-function $\beta={\partial a_s\over\partial\ln\mu^2}$ are known
up to four-loop order in
Refs.~\cite{Chetyrkin:1997dh,Vermaseren:1997fq,vanRitbergen:1997va,Czakon:2004bu}.
Indeed $\beta$, $\gamma_m$ and $\gamma_{VV}$ are known at least one
order higher in perturbation theory than needed in order to check the
$\mu$-dependent part of the vacuum polarization function up to four-loop
order.  This allows in turn to predict the $\mu$-dependent part of
$\overline{\Pi}^{0h}_{t}(q^2=0,\overline{m}_t)$ at five-loop order from
Eq.~(\ref{eq:anodim}).  The $\mu$-independent part, however, remains
unknown, but it is also not needed for the computation of the
contributions of the Higgs-boson partial decay width into two photons
which are considered in this work.

In the next section we will start with the computation of the complete
decoupling function $\zeta_{g\gamma}^2$ up to four-loop order.  The
latter is then used in order to determine the contributions to the
amplitude $A_t^{\infty}$ of the Higgs-boson decay into two photons.  A
detailed description of the renormalization procedure of
$\zeta_{g\gamma}^2$ can be found in Ref.~\cite{Chetyrkin:1997un}.  The
bare and renormalized decoupling function are related by the field
renormalization constants $Z_{ph}$ of the photon field $A^{\mu}$ in the
effective and full theory,
\begin{equation}
\zeta_{ph}\equiv\zeta_{g\gamma}^2={Z'_{ph}\over Z_{ph}}{\zeta^0_{g\gamma}}^{\!\!2}\,,
\end{equation}
where we have introduced the shorthand
$\zeta_{ph}\equiv\zeta_{g\gamma}^2$.  The renormalization constant
$Z_{ph}$ in QCD can be found in
Refs.~\cite{Gorishnii:1990vf,Chetyrkin:1996ez} up to four-loop order.
We define the perturbative expansion of the decoupling function
$\zeta_{ph}$ by
\begin{eqnarray}
\label{eq:zeta0}
\zeta_{ph}=1+{\bar{\alpha}^{(\nf)}(\mu)\over4\*\pi}\*\dR\*
             \sum_{k=0}\as^{k}\zeta_{ph}^{(k)}(\overline{m}_t,\mu)
                   \,,
\end{eqnarray}
with the dimension of the quark representation of the color group $\dR$.
For the $SU(3)$ color group we have $\dR=3$. The expansion coefficients
up to order $\as^2$ are known from the results of
Refs.~\cite{Chetyrkin:1996cf,Chetyrkin:1997un} and are given in
Appendix~\ref{app:Pi0}.  At four-loop order ($k=3$) the contribution
proportional to $Q_t^2$ can be obtained partially from the results of
Refs.~\cite{Chetyrkin:2004fq,Chetyrkin:2006xg} for $SU(3)$
explicitly. In the next section we will generalize them for an arbitrary
value of $N_c$ and augment them also by the still unknown contributions
proportional to $Q_{q_i}^2$ and $Q_{q_i}\*Q_t$.

\section{Calculation and results\label{sec:results}}
The non-singlet, four-loop contribution in perturbative QCD to the
photon vacuum polarization function at $q^2=0$ was first computed in
Refs.~\cite{Chetyrkin:2004fq,Chetyrkin:2006xg} for the $SU(3)$ color
group. In this work we will add in addition the afore mentioned terms
$\overline{\Pi}^{h}_{q_i}(q^2=0,\overline{m}_t)$ as well as the singlet
contributions $\Pi_{tq_{i}}^{0h}(q^2=0,m_t^0)$ which are given by the
diagrams in which the two photons couple to two different fermion loops.
They are distinguished from the other diagrams by the multiplicative
color factor of the symmetric structure constant
$\dabc=\mbox{Tr}\left[\{T^{a},T^{b}\}T^{c}\right]/\tr$ with
$a,b,c=1,\dots,8$. The symbol $\tr$ denotes the normalization of the
trace of the $SU(3)$ generators $T^{a}=\lambda^a/2$ in the fundamental
representation, which is conventionally chosen as 1/2.  The symbols
$\lambda^a$ are the Gell-Mann matrices.  In order to reconstruct the
general $SU(\nc)$ color structure of all diagrams which contribute to
the four-loop QCD corrections we generate them in the first step with
the program {\tt{QGRAF}}~\cite{Nogueira:1991ex}.  The arising four-loop
integrals are then evaluated at $q^2=0$ and mapped to the proper
notation which is needed for the subsequent reduction process with the
programs {\tt{q2e}} and {\tt{exp}}~\cite{Seidensticker:1999bb,
  Harlander:1997zb}.  This reduction to master integrals is performed
with a
{\tt{FORM}}~\cite{Vermaseren:2000nd,Vermaseren:2002rp,Tentyukov:2006ys}
based program which employs the traditional integration-by-parts
method~\cite{Chetyrkin:1981qh} in combination with Laporta's
algorithm~\cite{Laporta:1996mq,Laporta:2001dd}.  We use
{\tt{FERMAT}}~\cite{Lewis:rh} for the simplification of the rational
functions in the space-time dimension $d$ which arise as coefficient
functions of the loop integrals. The remaining dimensionally regulated
master integrals are known to sufficient high order in the
$\vep$-expansion~\cite{ Schroder:2005va,Chetyrkin:2006dh} with
$\vep=(4-d)/2$.  Analytic results for specific master integrals or
specific orders in the $\vep$-expansion have also been obtained in
Refs.~\cite{ Broadhurst:1991fi,Schroder:2005db,
  Chetyrkin:2004fq,Broadhurst:1996az,
  Laporta:2002pg,Kniehl:2005yc,Kniehl:2006bf,Kniehl:2006bg}.

The resulting renormalized decoupling function
$\zeta_{ph}^{(3)}(\overline{m}_t,\mu)$ can be decomposed into several
gauge-invariant contributions.  In a first step we separate the three
terms which are distinguished by different combinations of the charge
factors $Q_{q_i}$ and $Q_t$ depending on whether the external photons
couple to a massless quark or a massive top quark,
\begin{equation}
\label{eq:zetagammadecomp}
\zeta_{ph}^{(3)}(\overline{m}_t,\mu)=
       Q_t^2\*\zeta_{ph,t}^{(3)}(\overline{m}_t,\mu)
      +\sum_{i=1}^{\nl}Q_{q_i}^2\*\zeta_{ph,q}^{(3)}(\overline{m}_t,\mu)
      +\sum_{i=1}^{\nl}Q_t\*Q_{q_i}\*\zeta_{ph,tq}^{(3)}(\overline{m}_t,\mu)\,.
\end{equation}
The non-singlet diagrams can be again subdivided according to the number
of inserted closed fermion loops. The symbol $\nl$ labels in the
following the number of massless quarks in an inserted closed fermion
loop, whereas $\nh=1$ labels the insertion of a massive fermion loop
into the vacuum polarization function. The renormalization of the terms
proportional to $Q_t^2$ and $Q_t\*Q_{q_i}$ in
Eq.~(\ref{eq:zetagammadecomp}) is straightforward. The results read
\begin{eqnarray}
\label{eq:C3t}
\zeta_{ph,t}^{(3)}(\overline{m}_t,\mu)&\!\!=\!\!&
  \cf^3\*\left[
  {37441\over8640} 
+ {1024\over15}\*\A5 
+ {7676\over45}\*\A4 
- {3429\over40}\*\z5
+ {7549\over80}\*\z3 
- {58001\over32400}\*\pi^4 
- {128\over225}\*\logtwo^5 
\right.\nonumber\\&&\left.
+ {1919\over270}\*\logtwo^4 
+ {128\over135}\*\logtwo^3\*\pi^2 
- {1919\over270}\*\logtwo^2\*\pi^2 
+ {424\over675}\*\logtwo\*\pi^4 
+ {157\over32}\*\LmusdmMSbars 
       \right]
- \cf^2\*\ca\*\left[
  {707191\over103680} 
\right.\nonumber\\&&\left.
+ {1696\over15}\*\A5 
+ {4891\over30}\*\A4 
- {11807\over80}\*\z5
+ {868901\over8640}\*\z3 
- {153599\over86400}\*\pi^4 
- {212\over225}\*\logtwo^5 
+ {4891\over720}\*\logtwo^4 
\right.\nonumber\\&&\left.
+ {212\over135}\*\logtwo^3\*\pi^2 
- {4891\over720}\*\logtwo^2\*\pi^2 
+ {1517\over1350}\*\logtwo\*\pi^4 
+ {55\over32}\*\LmusdmMSbars^2 
+ \LmusdmMSbars\*\left({2303\over288} - {407\over96}\*\z3\right) 
        \right]
\nonumber\\&&
+ \cf\*\ca^2\*\left[
  {1163113\over373248} 
+ {592\over15}\*\A5 
+ {6997\over180}\*\A4 
- {23209\over480}\*\z5
+ {5837\over384}\*\z3 
- {182893\over518400}\*\pi^4 
\right.\nonumber\\&&\left.
- {74\over225}\*\logtwo^5 
+ {6997\over4320}\*\logtwo^4 
+ {74\over135}\*\logtwo^3\*\pi^2 
- {6997\over4320}\*\logtwo^2\*\pi^2 
+ {1093\over2700}\*\logtwo\*\pi^4 
+ {121\over432}\*\LmusdmMSbars^3 
+ {461\over432}\*\LmusdmMSbars^2 
\right.\nonumber\\&&\left.
+ \LmusdmMSbars\*\left({4415\over648} - {935\over192}\*\z3\right) 
             \right] 
+\nh\*\cf\*\tr\*\left[
\cf\*\left(
  {2261597\over259200} 
+ {3496\over45}\*\A4 
+ {123149\over2700}\*\z3
\right.\right.\nonumber\\&&\left.\left.
- {29737\over32400}\*\pi^4 
+ {437\over135}\*\logtwo^4 
- {437\over135}\*\logtwo^2\*\pi^2 
+ {\LmusdmMSbars^2\over2} 
+ \LmusdmMSbars\*\left({41\over144} - {13\over24}\*\z3\right) 
     \right) 
\right.\nonumber\\&&\left.
+ \ca\*\left(
  {20108987\over16329600} 
+ {22\over45}\*\A4 
- {5\over6}\*\z5
+ {79649\over75600}\*\z3 
- {\pi^4\over8100} 
+ {11\over540}\*\logtwo^4 
- {11\over540}\*\logtwo^2\*\pi^2 
\right.\right.\nonumber\\&&\left.\left.
- {11\over54}\*\LmusdmMSbars^3 
- {79\over216}\*\LmusdmMSbars^2 
- \LmusdmMSbars\*\left({1189\over1296} + {\z3\over32}\right) 
      \right)
           \right]
- \nl\*\cf\*\tr\*\left[
 \cf\*\left(
  {16507\over10368} 
+ {50\over9}\*\A4 
\right.\right.\nonumber\\&&\left.\left.
+ {1093\over432}\*\z3
- {919\over12960}\*\pi^4 
+ {25\over108}\*\logtwo^4 
- {25\over108}\*\logtwo^2\*\pi^2 
- {\LmusdmMSbars^2\over2} 
- \LmusdmMSbars\*\left({41\over144} - {13\over24}\*\z3\right) 
        \right)
\right.\nonumber\\&&\left.
- \ca\*\left(
  {137657\over93312} 
+ {25\over9}\*\A4 
+ {3343\over864}\*\z3
- {353\over5184}\*\pi^4 
+ {25\over216}\*\logtwo^4 
- {25\over216}\*\logtwo^2\*\pi^2 
- {11\over54}\*\LmusdmMSbars^3 
\right.\right.\nonumber\\&&\left.\left.
- {79\over216}\*\LmusdmMSbars^2 
- \LmusdmMSbars\*\left({817\over324} - {37\over48}\*\z3\right) 
      \right) 
              \right]
- \nh^2\*\cf\*\tr^2\*\left[
  {610843\over816480} 
- {661\over945}\*\z3
- {\LmusdmMSbars^3\over27} 
+ {\LmusdmMSbars^2\over27}
\right.\nonumber\\&&\left. 
+ \LmusdmMSbars\*\left({113\over324} - {7\over24}\*\z3\right) 
                  \right]
- \nh\*\nl\*\cf\*\tr^2\*\left[
  {7043\over11664} 
- {2\over3}\*\A4 
- {127\over108}\*\z3
+ {49\over4320}\*\pi^4 
- {\logtwo^4\over36} 
\right.\nonumber\\&&\left.
+ {\pi^2\over36}\*\logtwo^2 
- {2\over27}\*\LmusdmMSbars^3 
+ {2\over27}\*\LmusdmMSbars^2 
+ \LmusdmMSbars\*\left({37\over324} - {7\over24}\*\z3\right) 
                      \right]
- \nl^2\*\cf\*\tr^2\*\left[
  {17897\over23328} 
- {31\over54}\*\z3
\right.\nonumber\\&&\left.
- {\LmusdmMSbars^3\over27} 
+ {\LmusdmMSbars^2\over27} 
- {19\over81}\*\LmusdmMSbars 
                  \right]
- {\dabc\*\dabc\over\dR}\*\left[
  {2411\over20160} 
- {73\over24}\*\A4 
- {5\over48}\*\z5
- {6779\over4480}\*\z3 
\right.\nonumber\\&&\left.
+ {2189\over69120}\*\pi^4 
- {73\over576}\*\logtwo^4 
+ {73\over576}\*\logtwo^2\*\pi^2 
+ \LmusdmMSbars\*\left({11\over144} - {\z3\over6}\right) 
                      \right]\,,\\
\label{eq:C3qt}
\zeta_{ph,tq}^{(3)}(\overline{m}_t,\mu)&\!\!=\!\!&
-{\dabc\*\dabc\over\dR}\*\left[
   {103\over864} 
 - {\pi^4\over720} 
 + {131\over288}\*\z3 
 - {5\over24}\*\z5 
 + \left({11\over72} - {\z3\over3}\right)\*\LmusdmMSbars
          \right]
\,,
\end{eqnarray}
where the symbol $\A{n}=\mbox{Li}_n(1/2)$ is given by the polylogarithm
function $\mbox{Li}_n(z)=\sum_{k=1}^{\infty}{z^k/k^n}$ and the Riemann
zeta-function is $\z{n}=\mbox{Li}_n(1)$.  The color factors
$\cf={(\nc^2-1)/(2\*\nc)}$ and $\ca=\nc$ denote the Casimir operators of
the $SU(\nc)$ group in the fundamental and adjoint representation. For
$\nc=3$ the explicit values read $\cf=4/3$, $\ca=3$ and
$\dabc\dabc=40/3$.  The symbols $\ell_{\mu}$ and $\logtwo$ are given by
the natural logarithm
$\ell_{\mu}=\ln\left({\mu^2/\overline{m}_t^2}\right)$ and
$\logtwo=\ln(2)$. \\
The renormalization of the contribution which is proportional to
$Q_{q_i}^2$ in Eq.~(\ref{eq:zetagammadecomp}) requires in addition the
QCD decoupling function of the strong coupling constant including higher
orders in the $\vep$-expansion. The latter can be extracted from the
calculation of the QCD decoupling function of
Refs.~\cite{Chetyrkin:2005ia,Schroder:2005hy}; or, alternatively, they
can be derived by combining the results of
Refs.~\cite{Grozin:2007fh,Grozin:2012ic,Marquard:2007uj}.  The result
for $\zeta_{ph,q}^{(3)}(\overline{m}_t,\mu)$ reads
\begin{eqnarray}
\label{eq:C3qq}
\zeta_{ph,q}^{(3)}(\overline{m}_t,\mu)&\!\!=\!\!&
     \nh\*\cf\*\tr\*\left[\cf\*\left(
  {14075\over10368} 
- {16\over3}\*\A4 
- {2\over9}\*\logtwo^4 
+ {2\over9}\*\logtwo^2\*\pi^2 
+ {89\over2160}\*\pi^4 
- {431\over216}\*\z3 
+ \left({23\over432} 
\right.\right.\right.\nonumber\\&&\left.\left.\left.
- {11\over9}\*\z3\right)\*\LmusdmMSbars 
+ {5\over12}\*\LmusdmMSbars^2
                    \right) 
- \ca\*\left(
  {39959\over93312} 
- {8\over3}\*\A4 
- {\logtwo^4\over9} 
+ {\pi^2\over9}\*\logtwo^2 
+ {47\over2160}\*\pi^4 
- {13\over36}\*\z3 
\right.\right.\nonumber\\&&\left.\left.
+ \left({403\over972} - {11\over9}\*\z3\right)\*\LmusdmMSbars 
+ {17\over72}\*\LmusdmMSbars^2 
+ {11\over108}\*\LmusdmMSbars^3
                    \right)
              \right]
 + \nh^2\*\cf\*\tr^2\*\left[
   {5423\over23328} 
 - {25\over54}\*\z3 
 + {31\over81}\*\LmusdmMSbars 
\right.\nonumber\\&&\left.
 - {11\over108}\*\LmusdmMSbars^2 
 + {2\over27}\*\LmusdmMSbars^3
                    \right]
 - \nh\*\nl\*\cf\*\tr^2\*\left[
   {6625\over11664} 
 - {11\over27}\*\z3 
 - {449\over972}\*\LmusdmMSbars 
 - {\LmusdmMSbars^3\over27}
                        \right] 
\,.
\end{eqnarray}
We used the RGE in Eq.~(\ref{eq:anodim}) to check the $\mu$-dependent
part of the calculation. Eqs.~(\ref{eq:C3t})-(\ref{eq:C3qq}) extend the
result of Ref.~\cite{Chetyrkin:1997un} for the decoupling function
$\zeta_{ph}$ by one order in perturbation theory.

In the next step we exploit Eq.~(\ref{eq:Lhgg}) and (\ref{eq:LHgamgam})
respectively in order to derive in the heavy top-quark mass limit the
four-loop QCD corrections for the fermionic amplitude $A_t^{\infty}$ of
the decay $H\to\gamma\gamma$.  The perturbative expansion of the
$\MSbar$ renormalized amplitude $A_t^{\infty}$ is given by
\begin{eqnarray}
\label{eq:Atexp}
A_t^{\infty}&=&\hat{A}_{t}\*\left[
  1
+ \as  \*A^{(1)}
+ \as^2\*A^{(2)}
+ \as^3\*A^{(3)}
+ \as^4\*A^{(4)}
+ \mathcal{O}\left(\as^5\right)
                          \right]\,.
\end{eqnarray}
The expansion coefficients $A^{(1)}$ and $A^{(2)}$ are given for
completeness in Appendix~\ref{app:A1toA2}. As described already in the
introductory text, at one-, two- and three-loop order the results are
also known with the top-quark mass dependence. At four- and five-loops
we restrict ourselves to those contributions of $A^{(3)}$ and $A^{(4)}$
for which the {\it{photons}} couple to a {\it{massive top-quark}} loop
and which can be related to the vacuum polarization function
$\overline{\Pi}_{t}^{h}(q^2=0,\overline{m}_t)$.  These contributions
will be labeled in the following by $A^{(3)}|_{\mbox{\tiny{$\gamma
      tt$-approx}}}$ and $A^{(4)}|_{\mbox{\tiny{$\gamma tt$-approx}}}$.

\begin{figure}[ht!]
\begin{center}
\begin{minipage}{3cm}
\begin{center}
\includegraphics[bb=72 452 540 720,width=3cm]{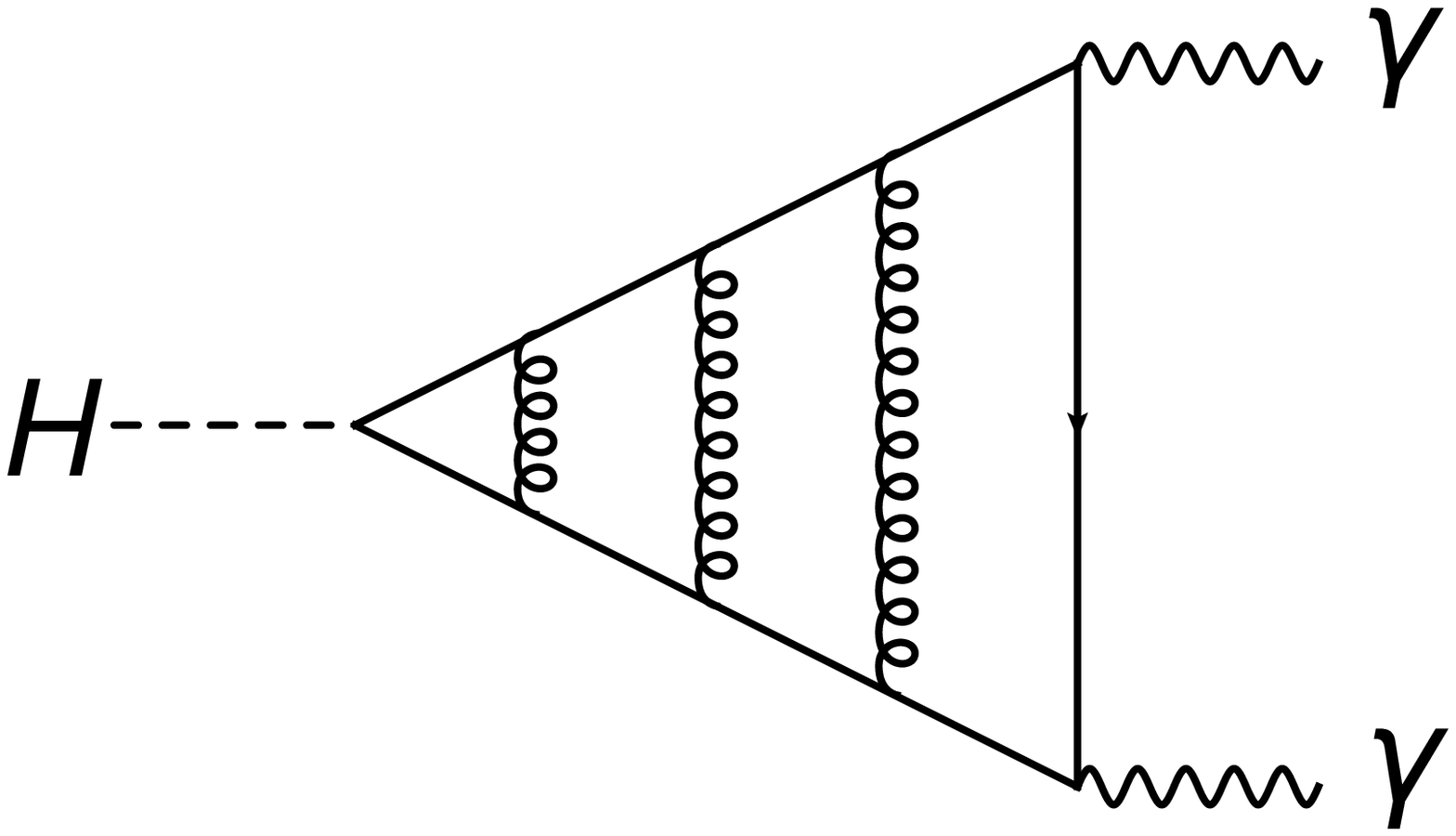}\\
$\cf^3$
\end{center}
\end{minipage}
\begin{minipage}{3cm}
\begin{center}
\includegraphics[bb=72 452 540 720,width=3cm]{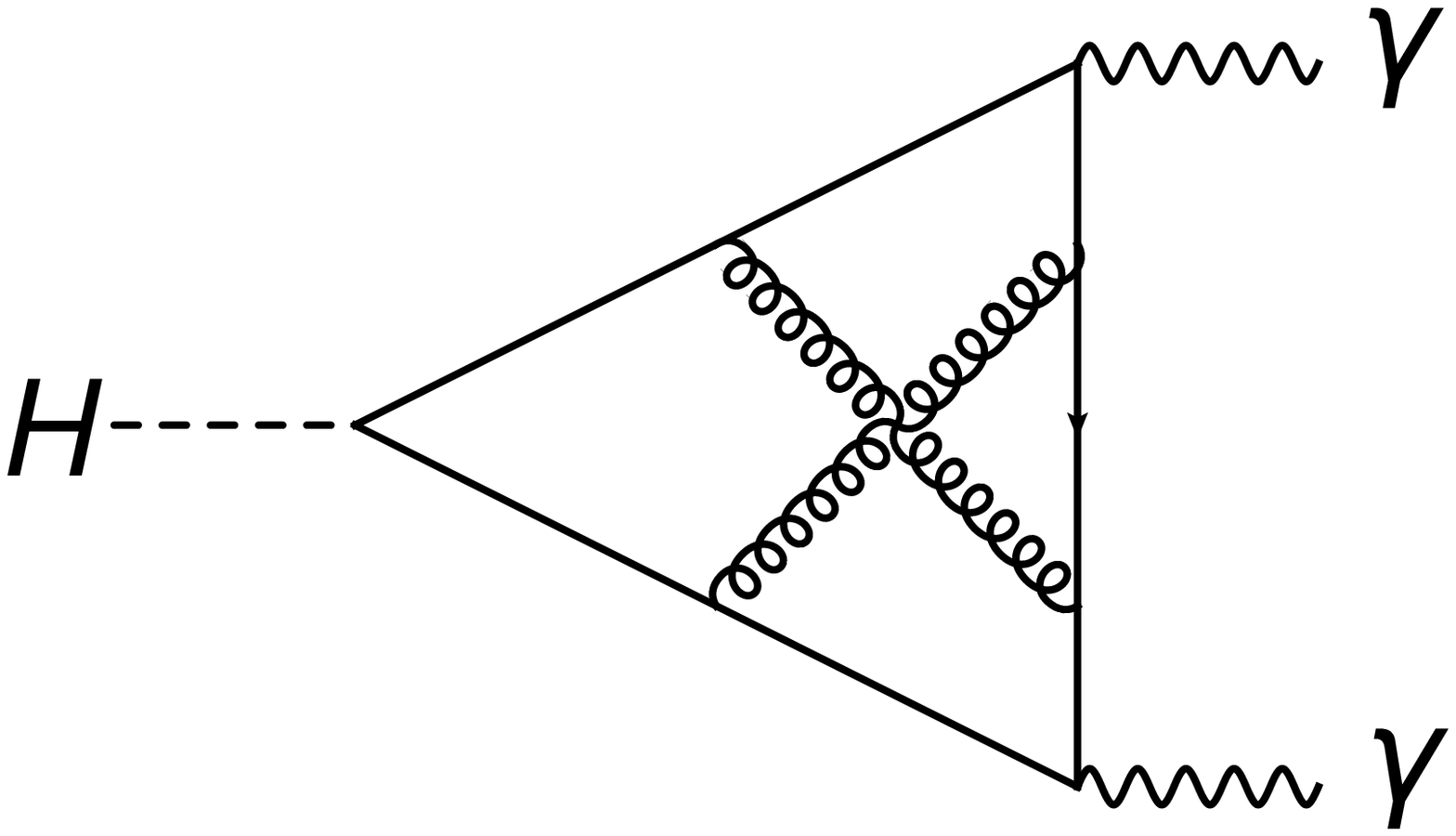}\\
$\cf^2\ca$
\end{center}
\end{minipage}
\begin{minipage}{3cm}
\begin{center}
\includegraphics[bb=72 452 540 720,width=3cm]{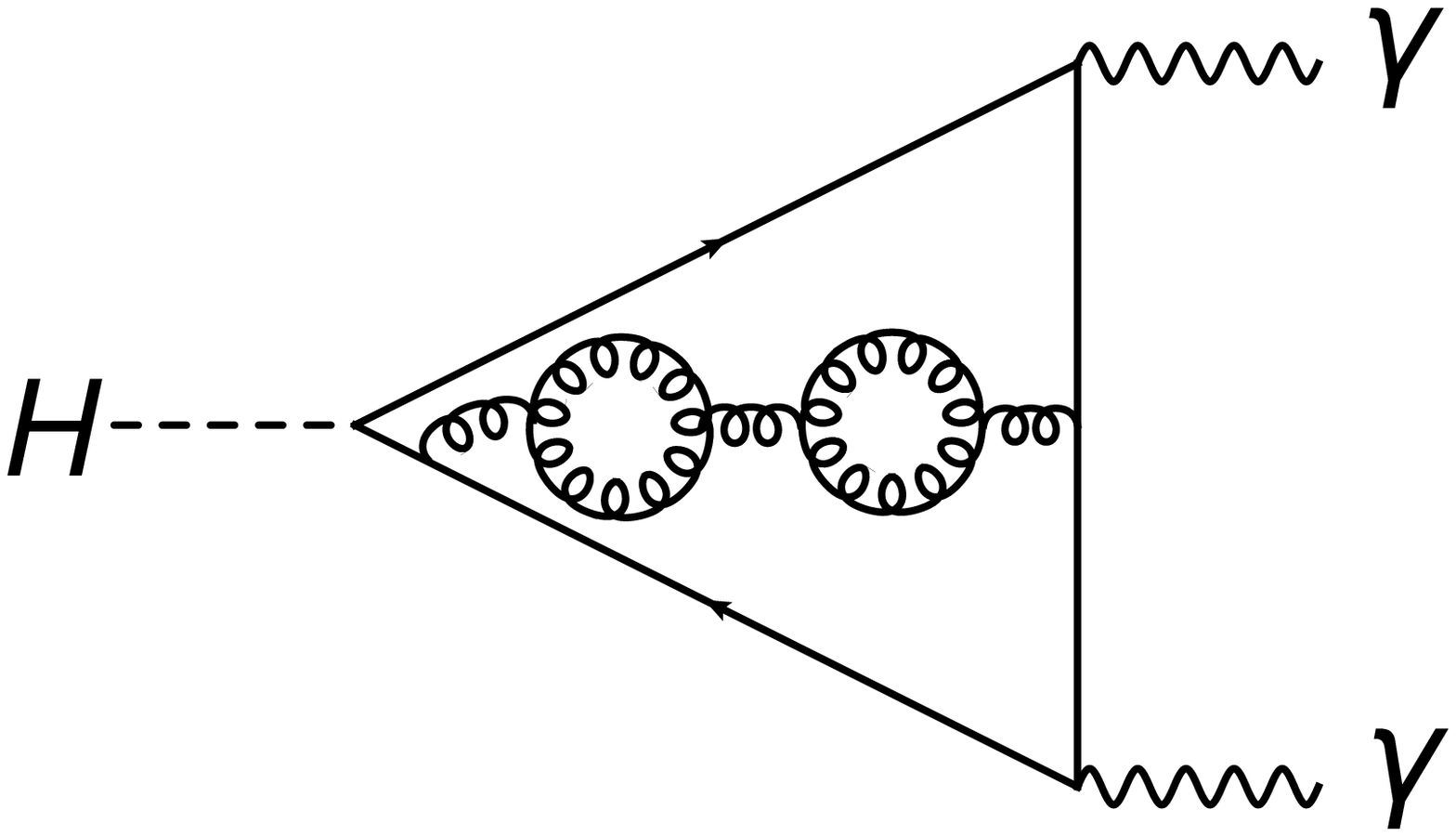}\\
$\cf\ca^2$
\end{center}
\end{minipage}
\begin{minipage}{3cm}
\begin{center}
\includegraphics[bb=72 452 540 720,width=3cm]{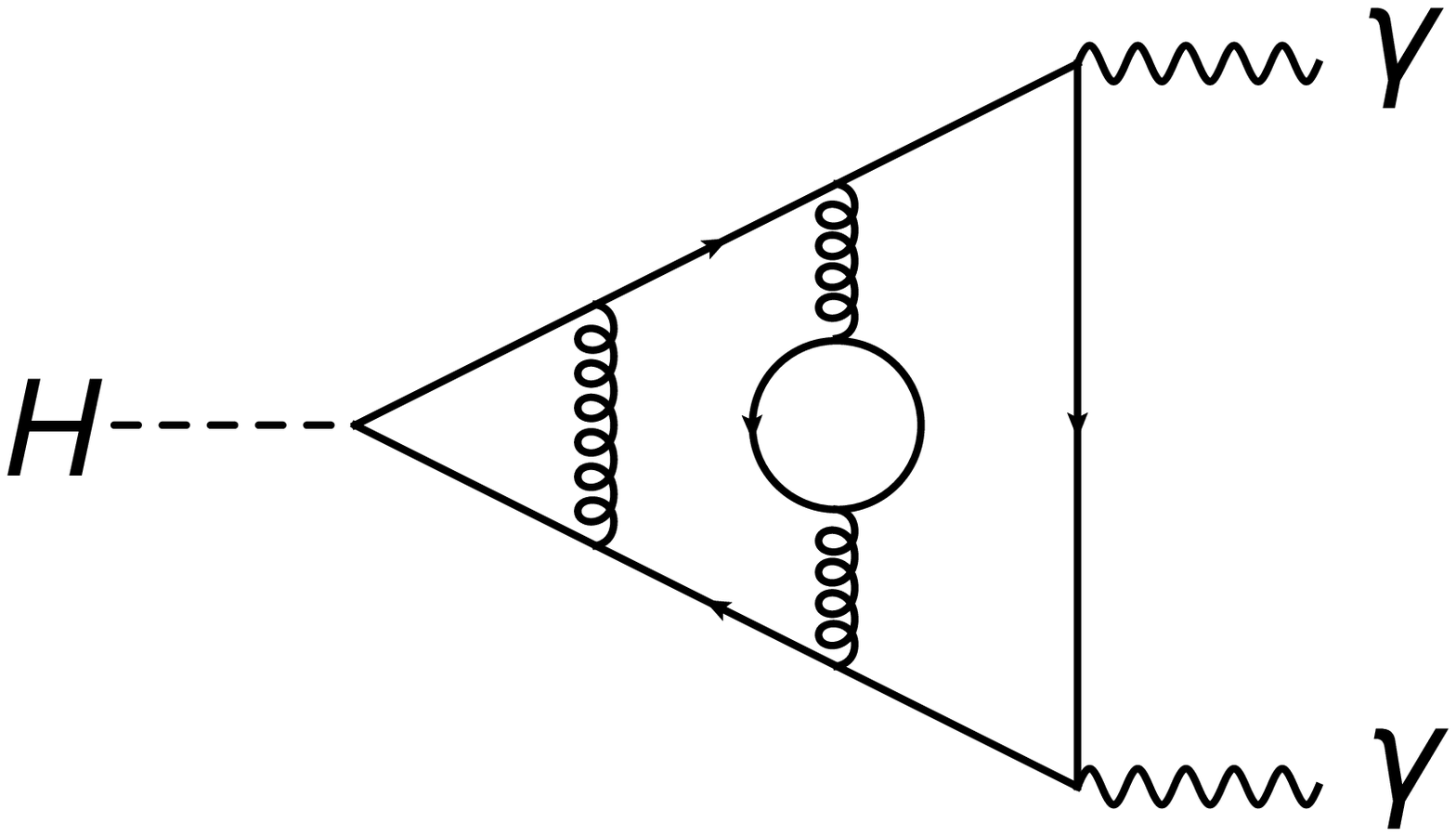}\\
$\nh\cf^2\tr$
\end{center}
\end{minipage}
\begin{minipage}{3cm}
\begin{center}
\includegraphics[bb=72 452 540 720,width=3cm]{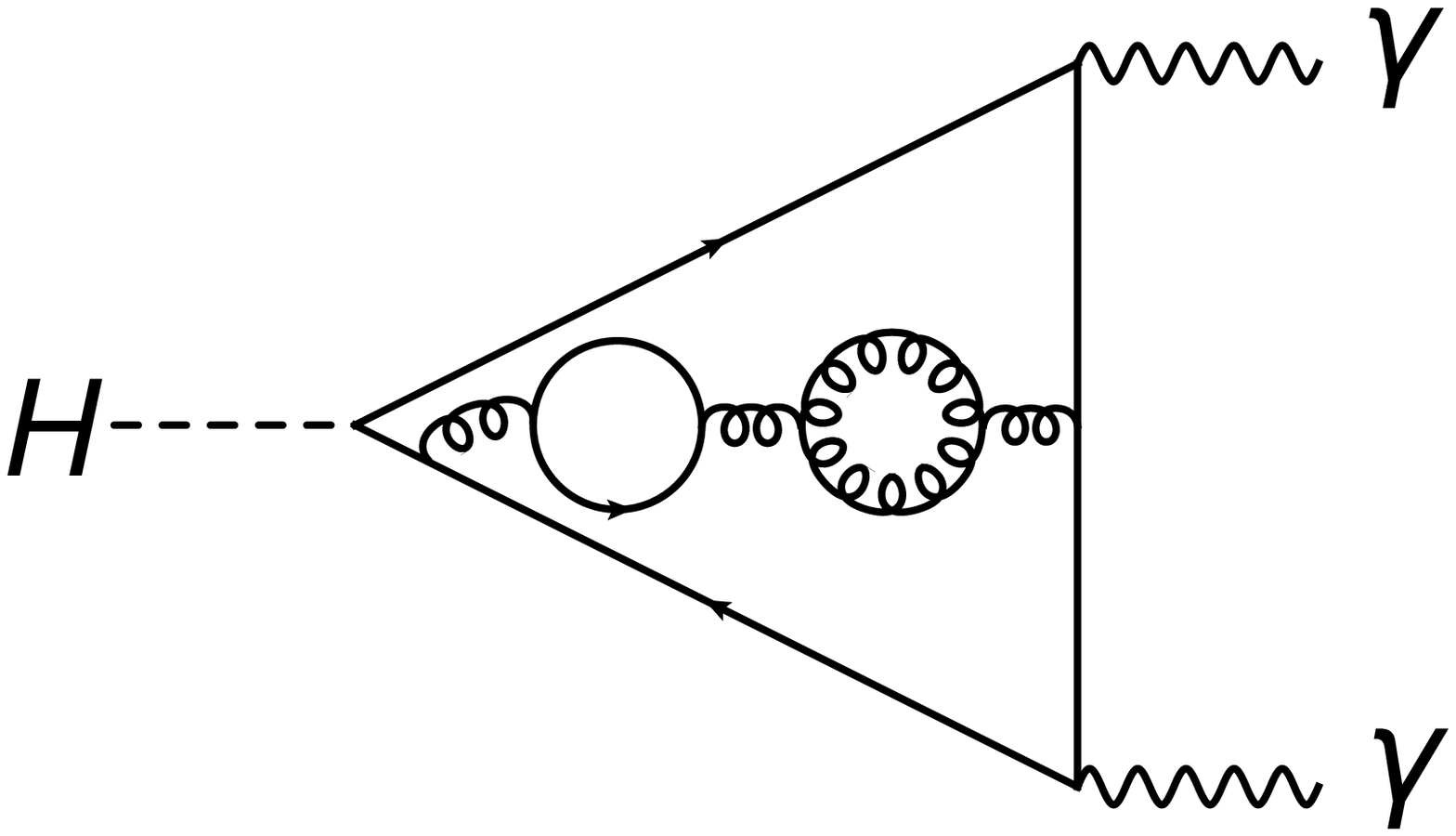}\\
$\nh\cf\ca\tr$
\end{center}
\end{minipage}\\[2ex]
\begin{minipage}{3cm}
\begin{center}
\includegraphics[bb=72 452 540 720,width=3cm]{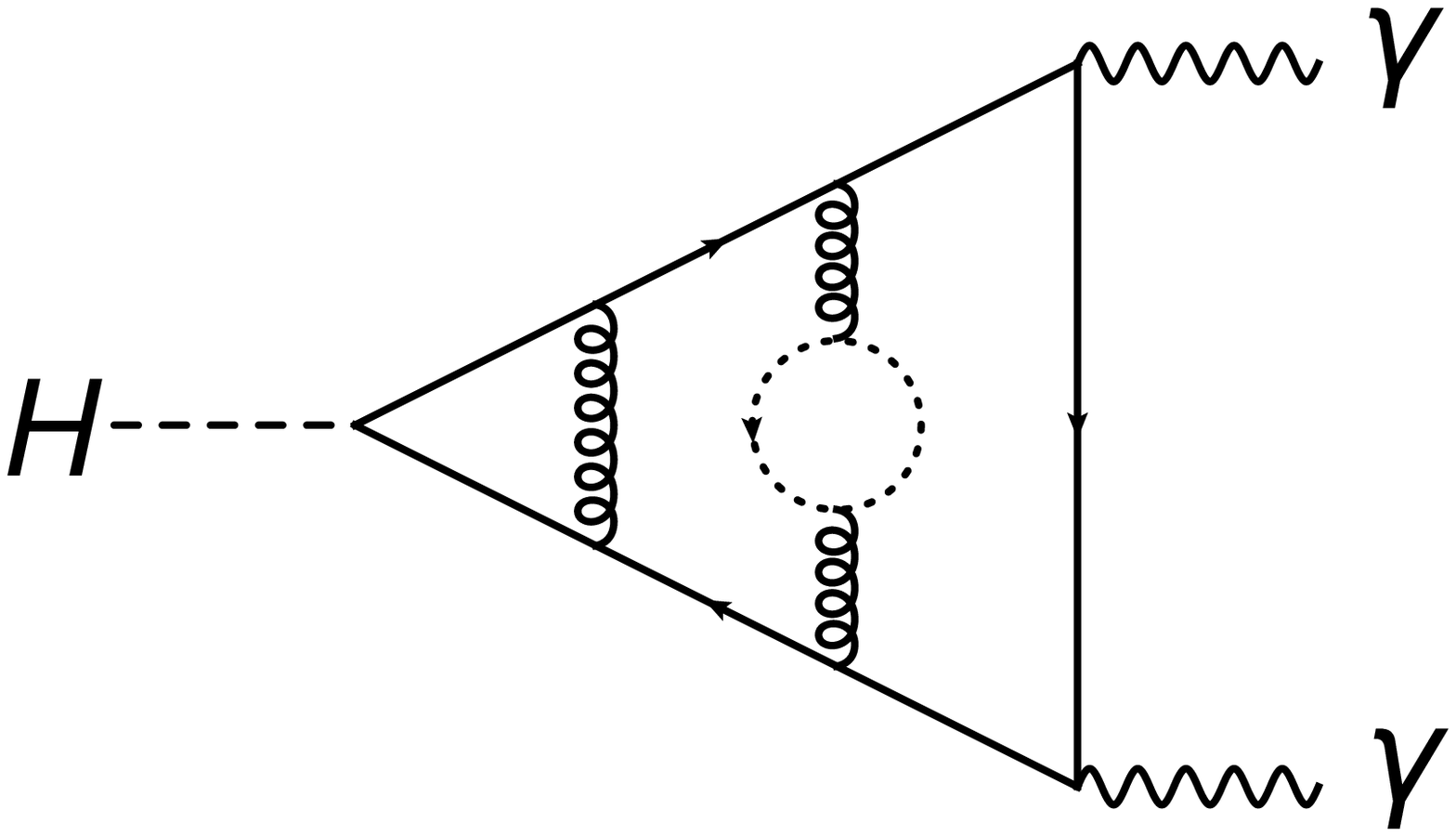}\\
$\nl\cf^2\tr$
\end{center}
\end{minipage}
\begin{minipage}{3cm}
\begin{center}
\includegraphics[bb=72 452 540 720,width=3cm]{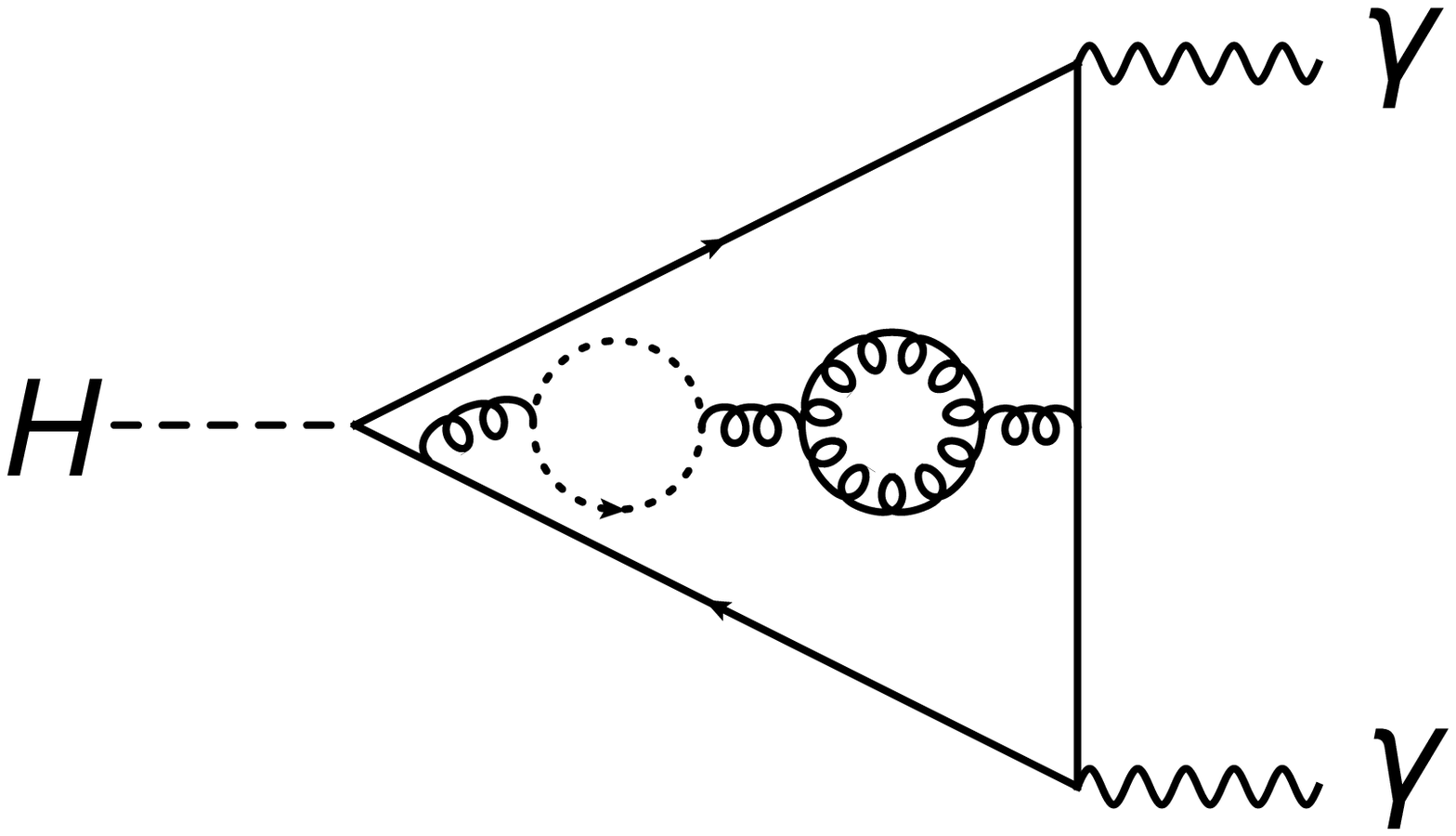}\\
$\nl\cf\ca\tr$
\end{center}
\end{minipage}
\begin{minipage}{3cm}
\begin{center}
\includegraphics[bb=72 452 540 720,width=3cm]{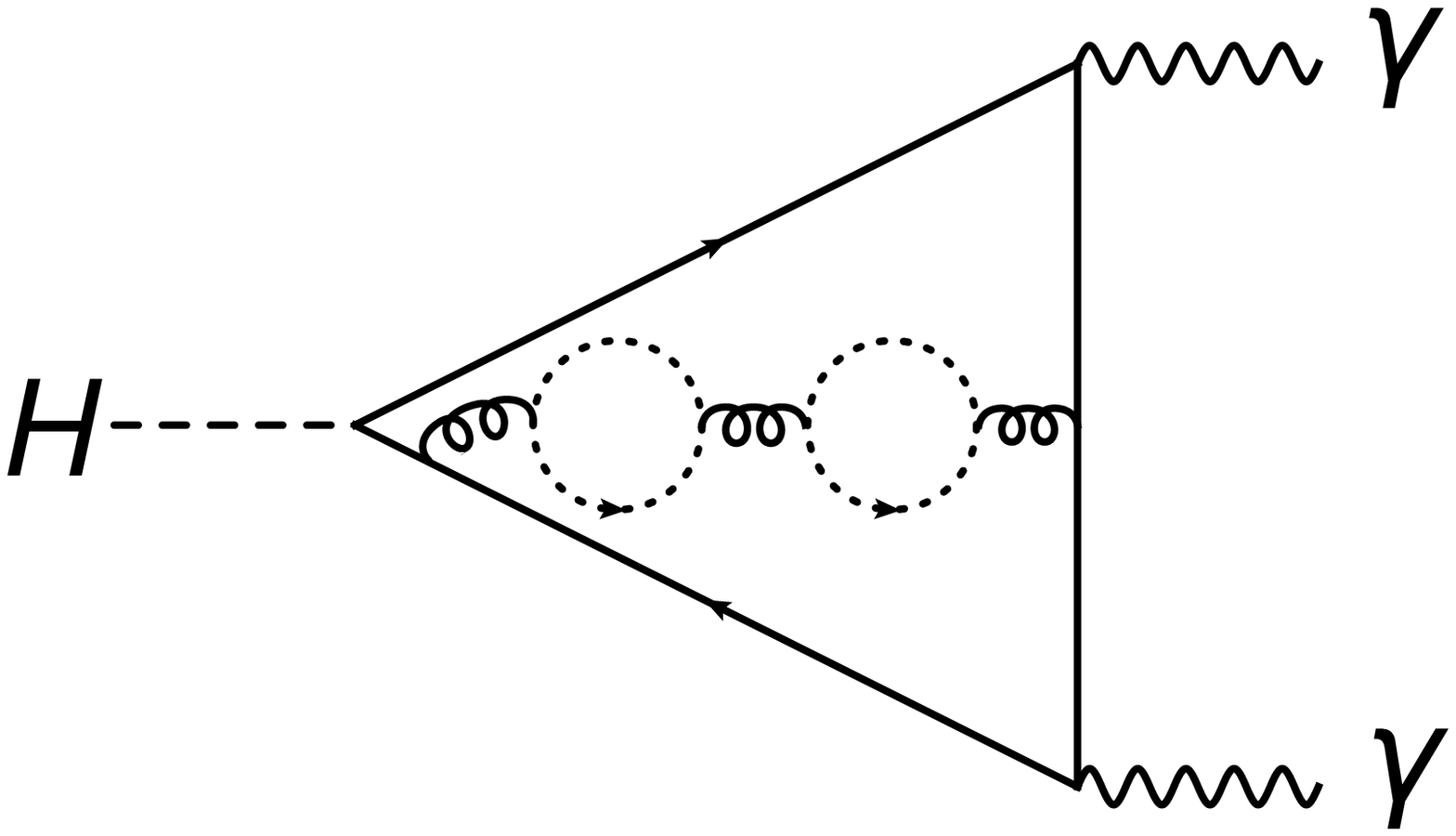}\\
$\nl^2\cf\tr^2$
\end{center}
\end{minipage}
\begin{minipage}{3cm}
\begin{center}
\includegraphics[bb=72 452 540 720,width=3cm]{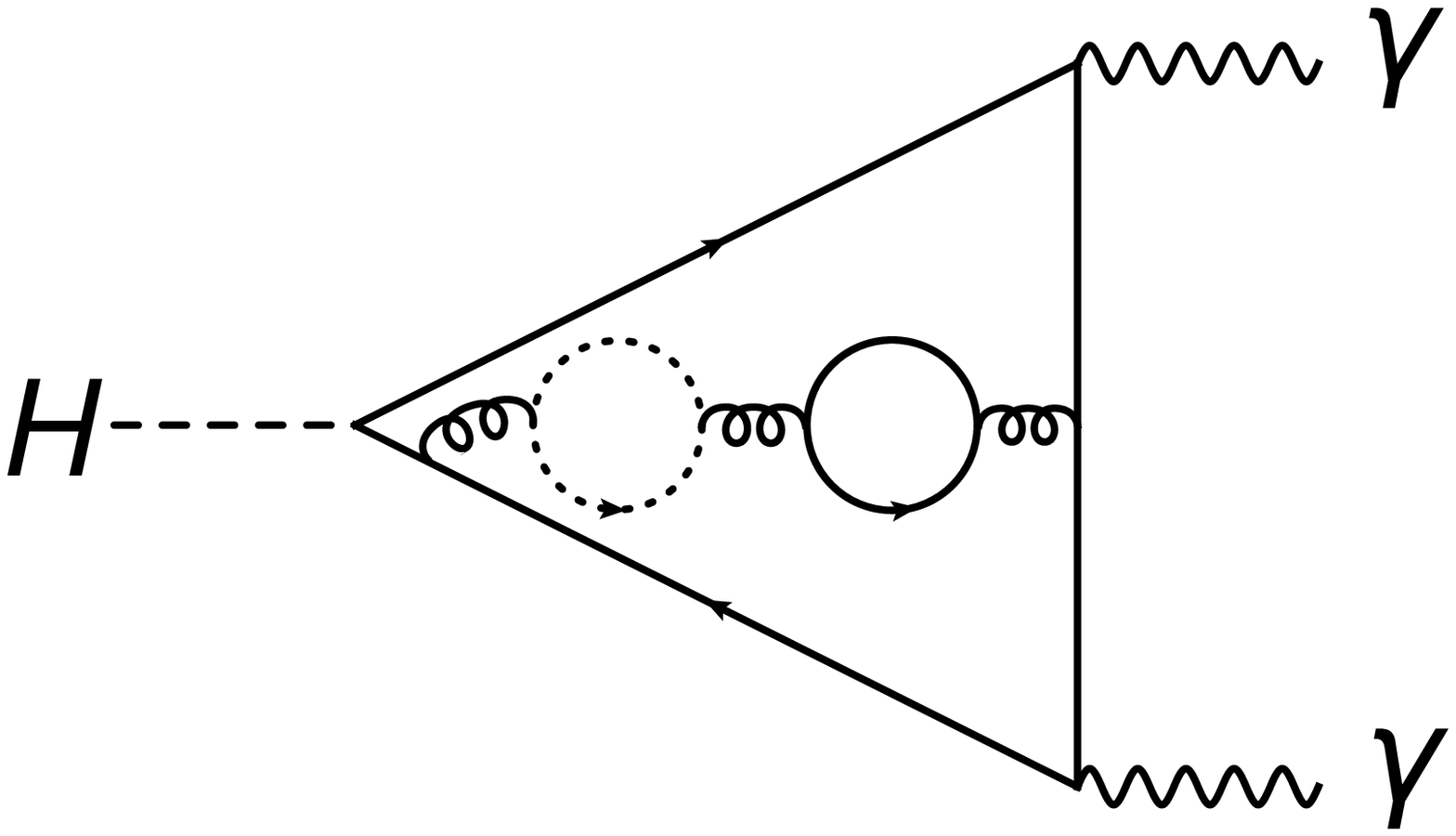}\\
$\nh\nl\cf\tr^2$
\end{center}
\end{minipage}
\begin{minipage}{3cm}
\begin{center}
\includegraphics[bb=72 452 540 720,width=3cm]{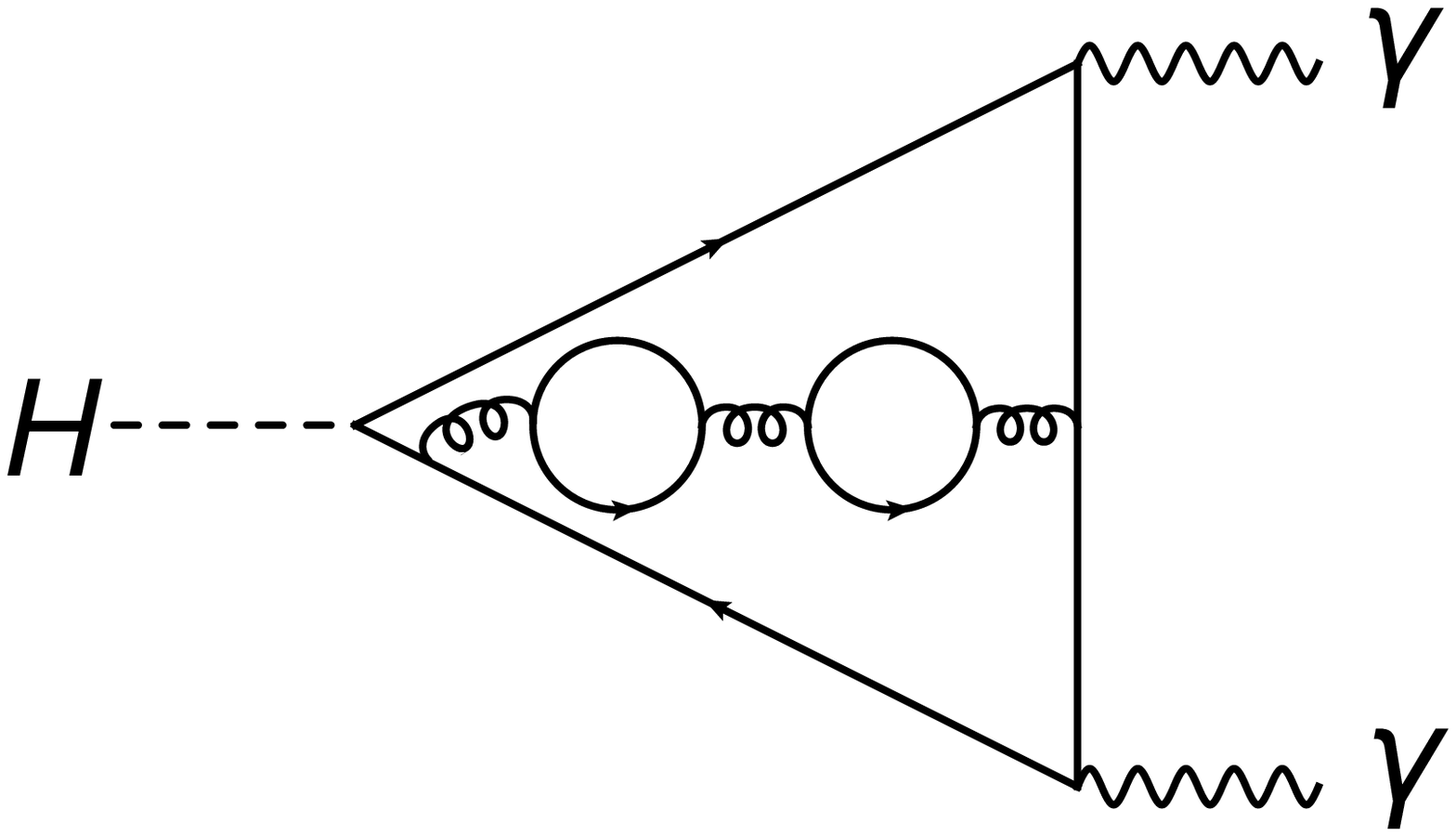}\\
$\nh^2\cf\tr^2$
\end{center}
\end{minipage}\\[2ex]
%
%
\begin{minipage}{3cm}
\begin{center}
\includegraphics[bb=72 452 540 720,width=3cm]{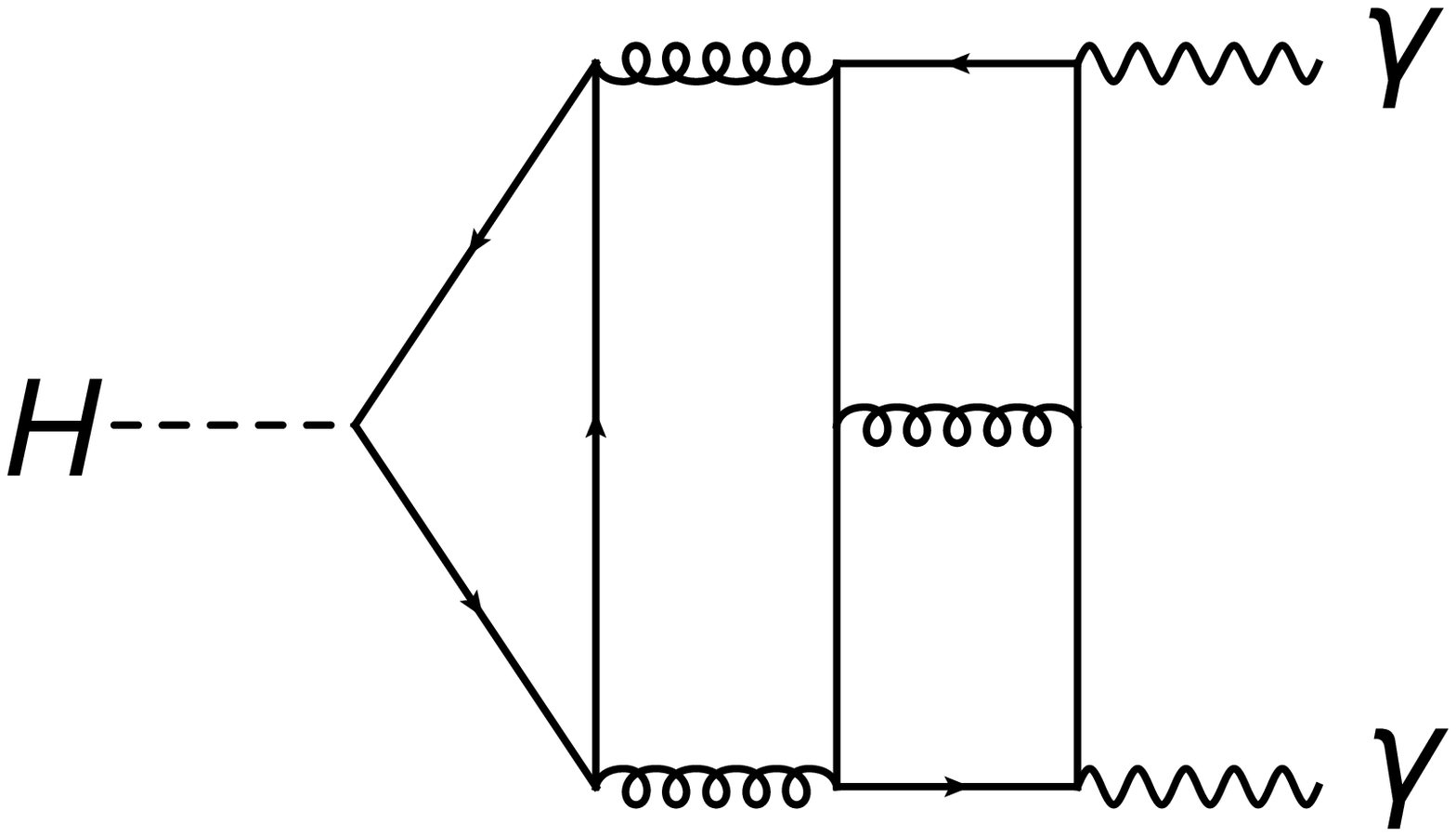}\\
$\si\,\nh\cf^2\tr$
\end{center}
\end{minipage}
\begin{minipage}{3cm}
\begin{center}
\includegraphics[bb=72 452 540 720,width=3cm]{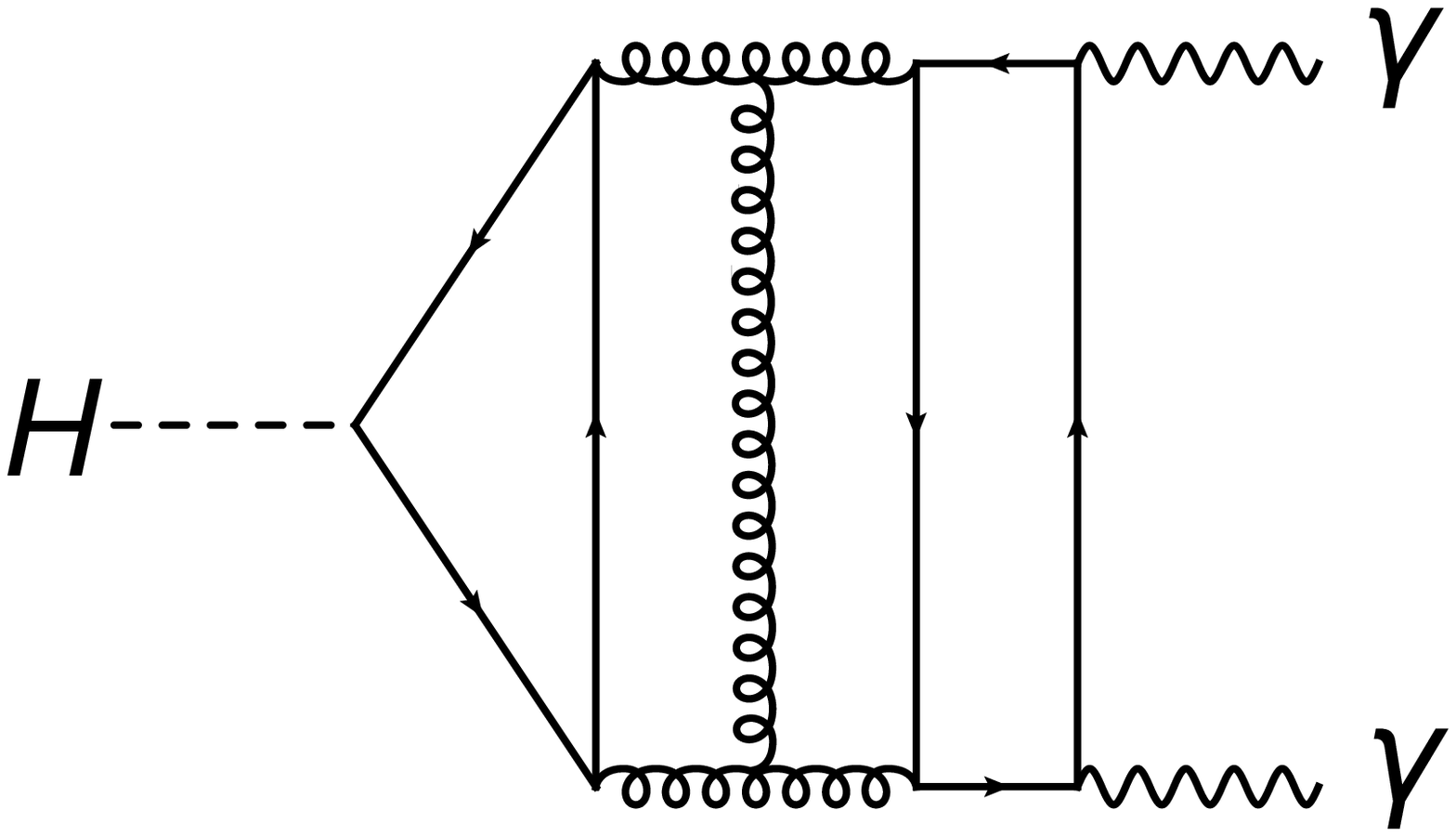}\\
$\si\,\nh\cf\ca\tr$
\end{center}
\end{minipage}
\begin{minipage}{3cm}
\begin{center}
\includegraphics[bb=76 359 539 622,width=3cm]{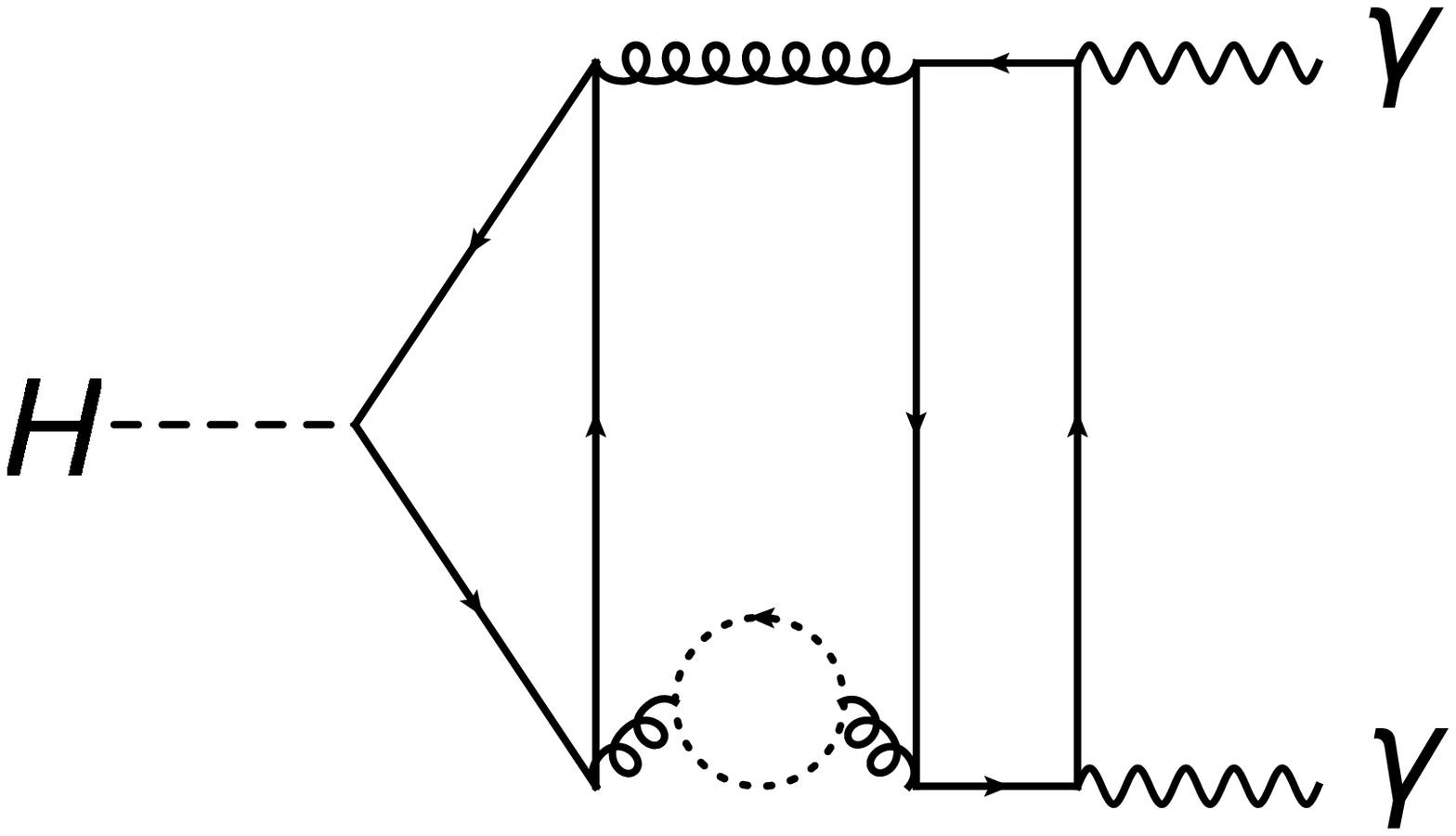}\\
$\si\,\nh\nl\cf\tr^2$
\end{center}
\end{minipage}
\begin{minipage}{3cm}
\begin{center}
\includegraphics[bb=76 359 539 622,width=3cm]{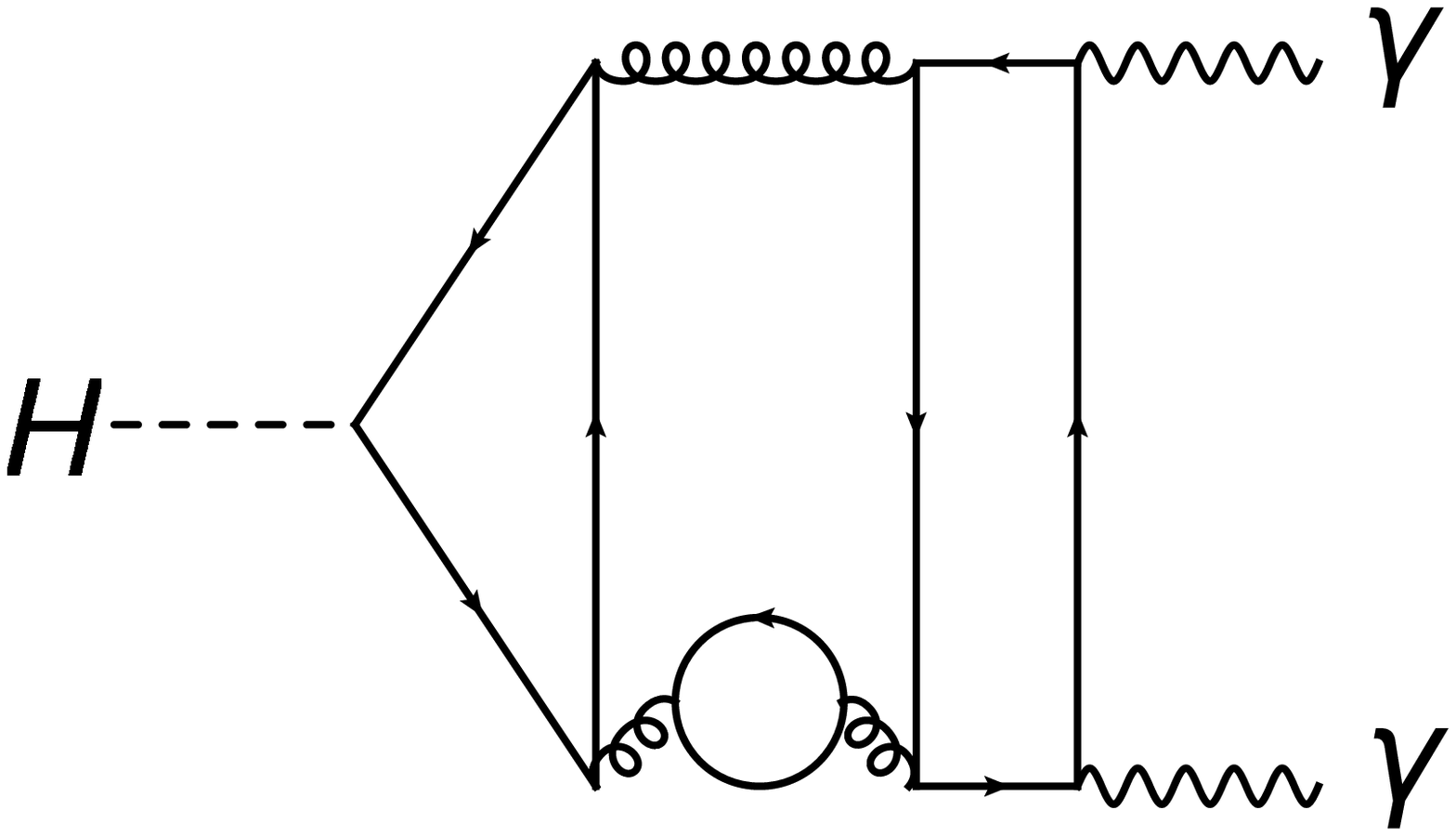}\\
$\si\,\nh^2\cf\tr^2$
\end{center}
\end{minipage}
\begin{minipage}{3cm}
\begin{center}
\includegraphics[bb=72 433 540 720,width=3cm]{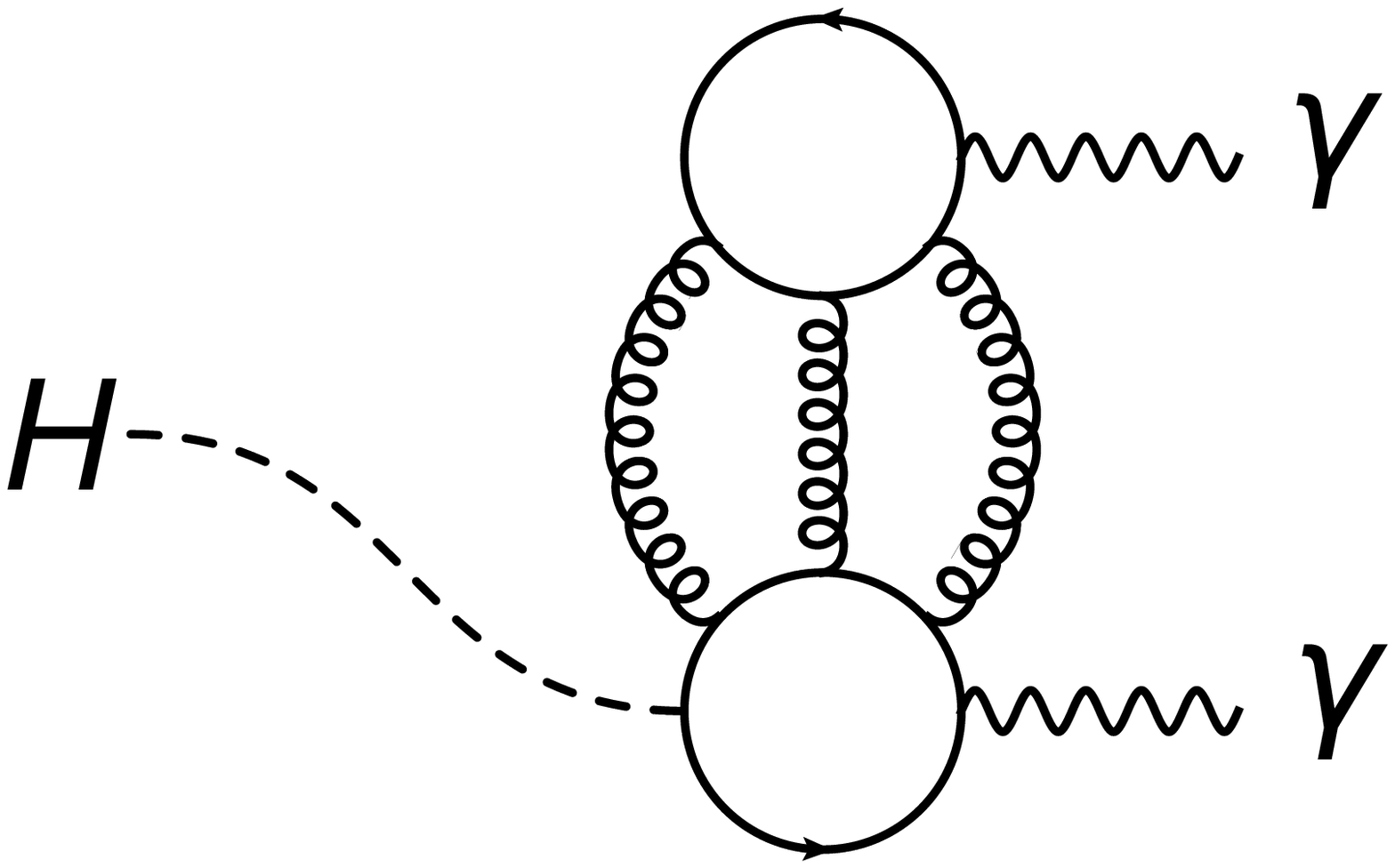}\\
$\nh\dabc\dabc$
\end{center}
\end{minipage}\\[2ex]
\end{center}
\vspace*{-0.4cm}
\caption{Example diagrams which illustrate the different kind of diagram
  classes which contribute to Eq.~(\ref{eq:A3}). Solid lines represent
  top quarks, dotted lines denote massless quarks, wavy lines are
  photons, twisted lines represent gluons and
  the dashed line is the Higgs boson.  The structure of the color and
  the flavor insertions of each diagram class is given below each
  example Feynman graph.
  \label{fig:4loop}}
\end{figure}

Starting from three-loop order in QCD there arise diagrams, where the
Higgs boson and the two photons do not couple to the same fermion
loop. In order to separate these singlet from the non-singlet
contributions, e.g. those diagrams where all three external particles
couple to the same fermion loop, we perform a second calculation. For
this purpose we act with the derivative with respect to the top-quark
mass on the amplitude already at the level of the diagram generation,
before the integration is performed and we introduce the label $\si=1$,
which serves only as a separator in order to distinguish these singlet
contributions from the remaining amplitude.  For each appearing color
structure which contributes to the amplitude
$A^{(3)}|_{\mbox{\tiny{$\gamma tt$-approx}}}$ of Eq.~(\ref{eq:Atexp}) we
show one example diagram for the whole diagram class in
Fig.~\ref{fig:4loop}.

The corresponding four-loop contribution reads
\begin{eqnarray}
\label{eq:A3}
\left.A^{(3)}\right|_{\mbox{\tiny{$\gamma tt$-approx}}}\!\!\!\!\!\!\!\!\!\!\!\!\!\!\!\!\!&=&
- \cf^3\*{471\over128} 
+ \cf^2\*\ca\*\left(
  {2303\over384} 
- {407\over128}\*\z3
+ {165\over64}\*\LmusdmMSbars 
             \right) 
- \cf\*\ca^2\*\left(
  {4415\over864} 
- {935\over256}\*\z3 
+ {121\over192}\*\LmusdmMSbars^2
\right.\nonumber\\&&\left.
+ {461\over288}\*\LmusdmMSbars 
            \right)
- \nh\*\cf\*\tr\*\left[
  \cf\*\left(
     {15589\over1600}
   + {376\over5}\*\A4 
   + {8681\over200}\*\z3 
   - {202\over225}\*\pi^4 
   + {47\over15}\*\logtwo^4
   - {47\over15}\*\logtwo^2\*\pi^2 
\right.\right.\nonumber\\&&\left.\left.
   + {3\over4}\*\LmusdmMSbars
       \right) 
+ \ca\*\left(
     {2109593\over1209600} 
   + {617\over80}\*\A4 
   - {5\over8}\*\z5 
   + {166463\over44800}\*\z3 
   - {19421\over230400}\*\pi^4 
   + {617\over1920}\*\logtwo^4 
\right.\right.\nonumber\\&&\left.\left.
   - {617\over1920}\*\logtwo^2\*\pi^2 
   - {11\over24}\*\LmusdmMSbars^2
   - {59\over288}\*\LmusdmMSbars
      \right)
               \right]
- \nl\*\cf\*\tr\*\left[
    \cf\*\left(
  {41\over192} 
- {13\over32}\*\z3 
+ {3\over4}\*\LmusdmMSbars
        \right) 
\right.\nonumber\\&&\left.
  - \ca\*\left(
  {817\over432} 
- {37\over64}\*\z3 
+ {11\over24}\*\LmusdmMSbars^2
+ {79\over144}\*\LmusdmMSbars 
        \right)
                \right] 
+ \nh^2\*\cf\*\tr^2\*\left(
  {767\over756} 
- {55\over56}\*\z3 
- {\LmusdmMSbars^2\over12}
+ {13\over72}\*\LmusdmMSbars 
                   \right) 
\nonumber\\&&
+ \nh\*\nl\*\cf\*\tr^2\*\left(
  {1021\over1728} 
- {91\over128}\*\z3 
- {\LmusdmMSbars^2\over6} 
+ {17\over72}\*\LmusdmMSbars
                      \right) 
- \nl^2\*\cf\*\tr^2\*\left(
  {19\over108} 
+ {\LmusdmMSbars^2\over12}
- {\LmusdmMSbars\over18} 
                   \right)
\nonumber\\&&
+\si\,\*\nh\*\cf\*\tr\*\left[
   \cf\*\left(
  {22871\over2400} 
+ {376\over5}\*\A4 
+ {35049\over800}\*\z3
- {202\over225}\*\pi^4 
+ {47\over15}\*\logtwo^4 
- {47\over15}\*\logtwo^2\*\pi^2 
      \right) 
\right.\nonumber\\&&\left.
+ \ca\*\left(
  {108959\over44800} 
+ {617\over80}\*\A4 
- {5\over8}\*\z5 
+ {167513\over44800}\*\z3 
- {19421\over230400}\*\pi^4 
+ {617\over1920}\*\logtwo^4 
- {617\over1920}\*\logtwo^2\*\pi^2 
\right.\right.\nonumber\\&&\left.\left.
+ {11\over32}\*\LmusdmMSbars
     \right)
- \nh\*\tr\*\left(
  {253\over336} 
- {171\over224}\*\z3 
+ {\LmusdmMSbars\over8}
           \right) 
- \nl\*\tr\*\left(
  {97\over192} 
- {63\over128}\*\z3 
+ {\LmusdmMSbars\over8}
           \right) 
   \right]
\nonumber\\&&
+ {\dabc\*\dabc\over\dR}\*\left({11\over192} - {\z3\over8}\right)\\
&\!\!\!\!\stackbin[\nc=3]{\mbox{\scriptsize{si}}=1=\nh}{=}\!\!\!\!&
- {95339\over2592} 
+ {7835\over288}\*\z3
- {961\over144}\*\LmusdmMSbars^2 
- {541\over108}\*\LmusdmMSbars 
+ \nl\*\left(
  {4693\over1296} 
- {125\over144}\*\z3 
+ {31\over36}\*\LmusdmMSbars^2
+ {101\over216}\*\LmusdmMSbars 
      \right) 
\nonumber\\&&
- \nl^2\*\left(
  {19\over324} 
+ {\LmusdmMSbars^2\over36}
- {\LmusdmMSbars\over54} 
        \right) 
+ \left[
  {55\over216} 
- {5\over9}\*\z3
 \right]
\,.
\label{eq:A3Nc3}
\end{eqnarray}
The result for $\nc=3$ of Eq.~(\ref{eq:A3Nc3}) agrees with the one of
Eq.~(55) in Ref.~\cite{Chetyrkin:1997un} which, in contrast, has been
obtained with the help of the mass anomalous dimension and
$\beta$-functions.  The last two terms in the square brackets of
Eq.~(\ref{eq:A3Nc3}) arise from the diagrams with the color structure
$\dabc\dabc$ in Eq.~(\ref{eq:A3}). They originate from the last diagram
class shown in Fig.~\ref{fig:4loop}. This separation allows for a
straightforward comparison with the results of
Ref.~\cite{Chetyrkin:1997un}.

At three-loop order mass corrections have been computed in
Refs.~\cite{Steinhauser:1996wy,Maierhofer:2012vv}. Since the mass of the
Higgs boson is now known to be $m_H\approx 126$~GeV one can expect that
these mass corrections are small, since they are suppressed by factors
of $\tau_t=m_H^2/(4\*\overline{m}_t^2)\approx0.14$ for
$\overline{m}_t(m_H)=166.79$~GeV. The latter value was derived from the
top-quark mass of $173.07$~GeV of Ref.~\cite{Beringer:1900zz}. For the
RGE running to different energy scales we use here and in the following
the program {\tt{RunDec}}~\cite{Chetyrkin:2000yt}. The smallness of the
mass corrections for the Higgs-boson mass of $m_H\approx 126$~GeV can
also be seen at three-loop order in Fig.~2 of
Ref.~\cite{Steinhauser:1996wy} as well as in Fig.~3 of
Ref.~\cite{Maierhofer:2012vv}. As a result of this we focus in this work
only on the lowest expansion coefficient.

\begin{figure}[ht!]
\begin{center}
\begin{minipage}{3cm}
\begin{center}
\includegraphics[bb=72 452 540 720,width=3cm]{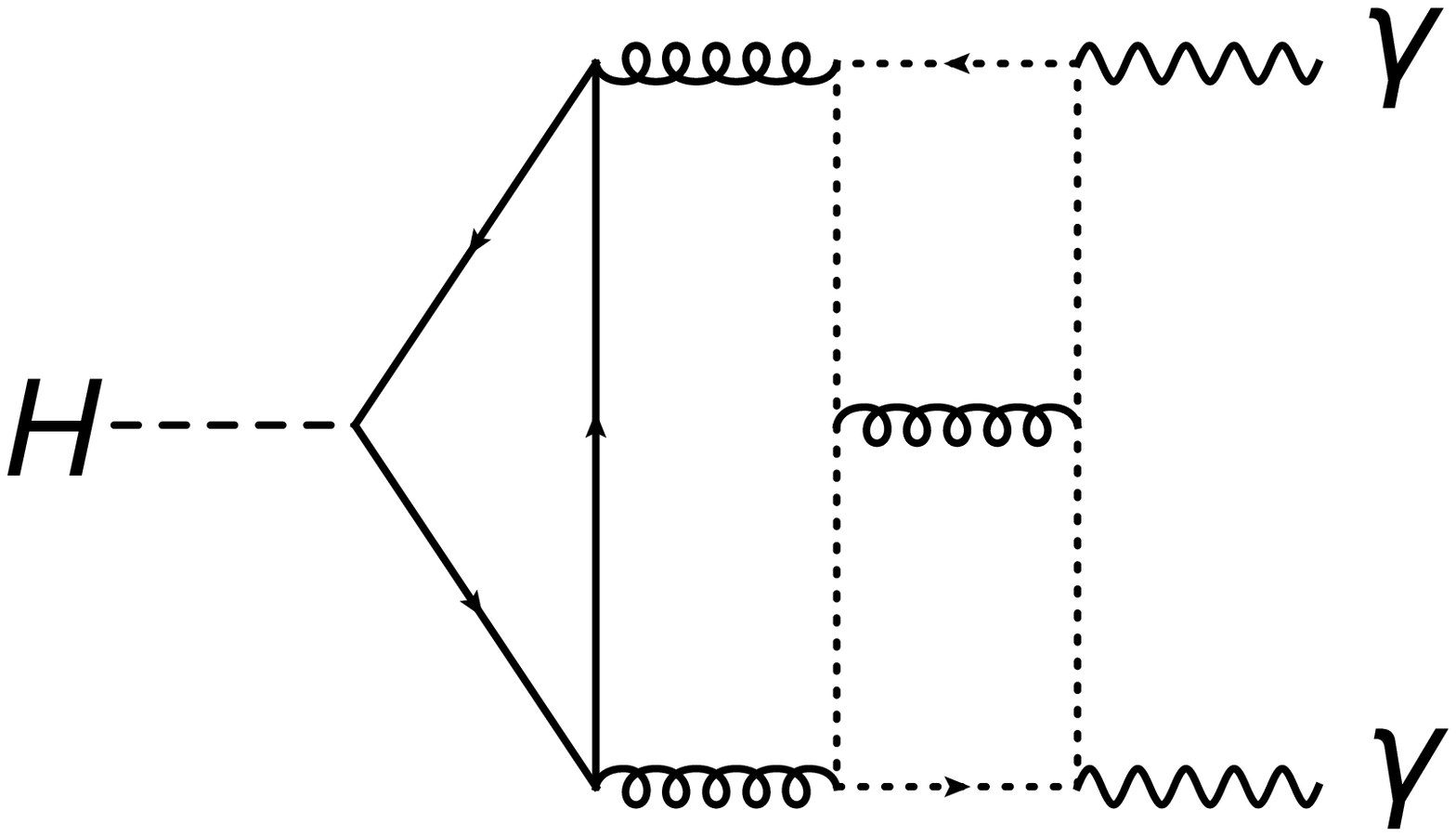}\\
$\si\,\nl\cf^2\tr$
\end{center}
\end{minipage}
\begin{minipage}{3cm}
\begin{center}
\includegraphics[bb=72 452 540 720,width=3cm]{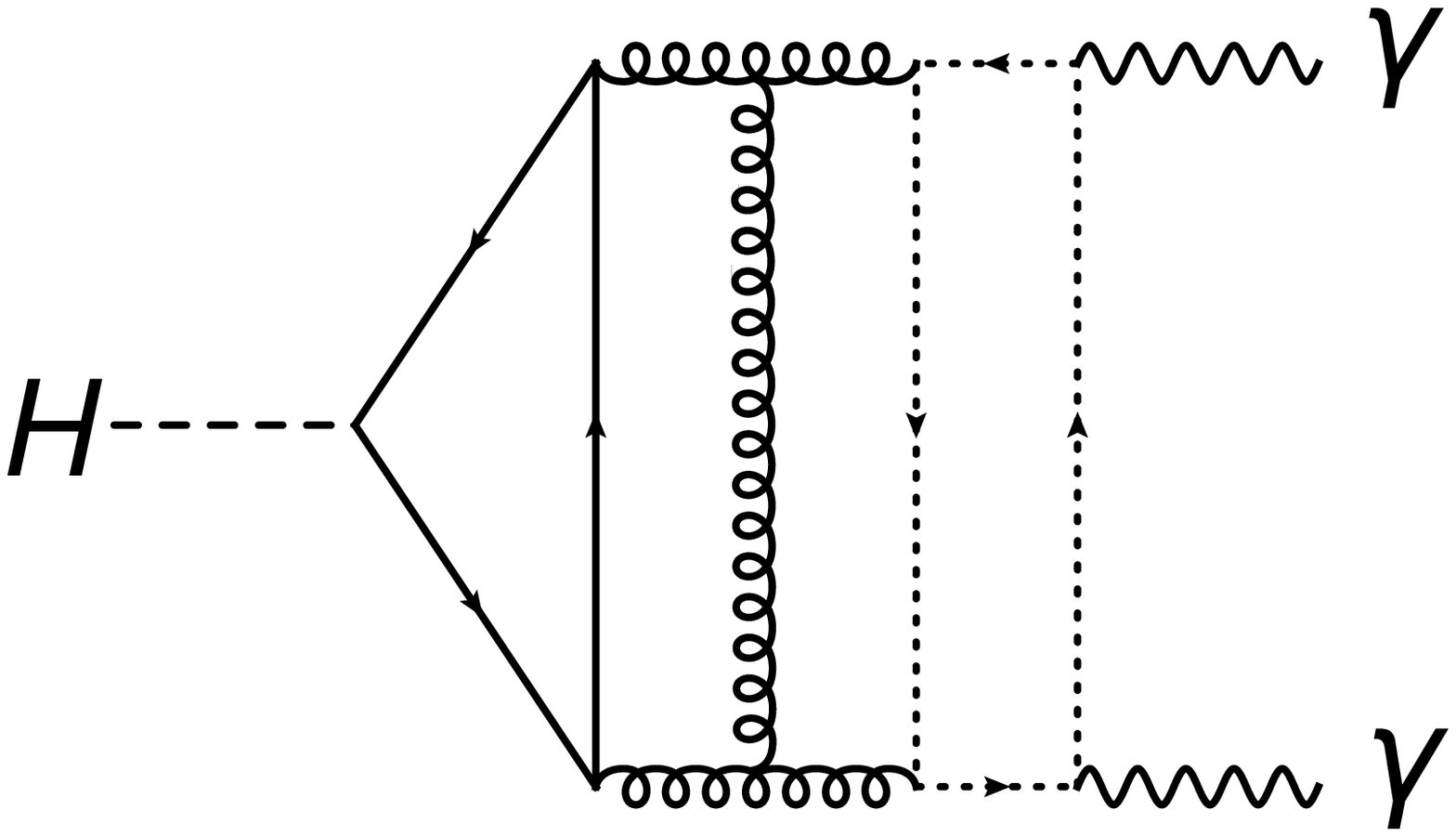}\\
$\si\,\nl\cf\ca\tr$
\end{center}
\end{minipage}
\begin{minipage}{3cm}
\begin{center}
\includegraphics[bb=76 359 539 622,width=3cm]{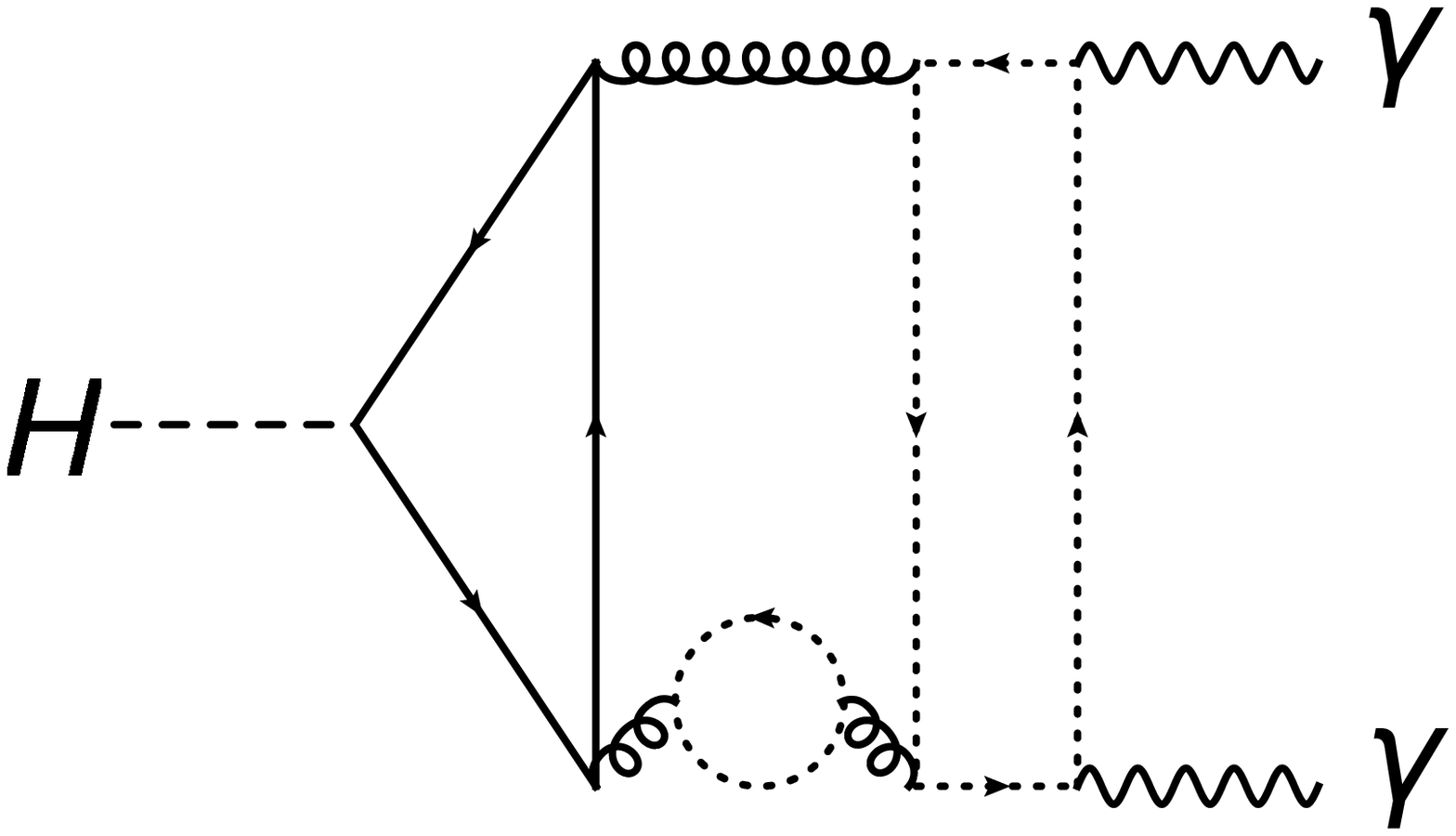}\\
$\si\,\nl^2\cf\tr^2$
\end{center}
\end{minipage}
\begin{minipage}{3cm}
\begin{center}
\includegraphics[bb=76 359 539 622,width=3cm]{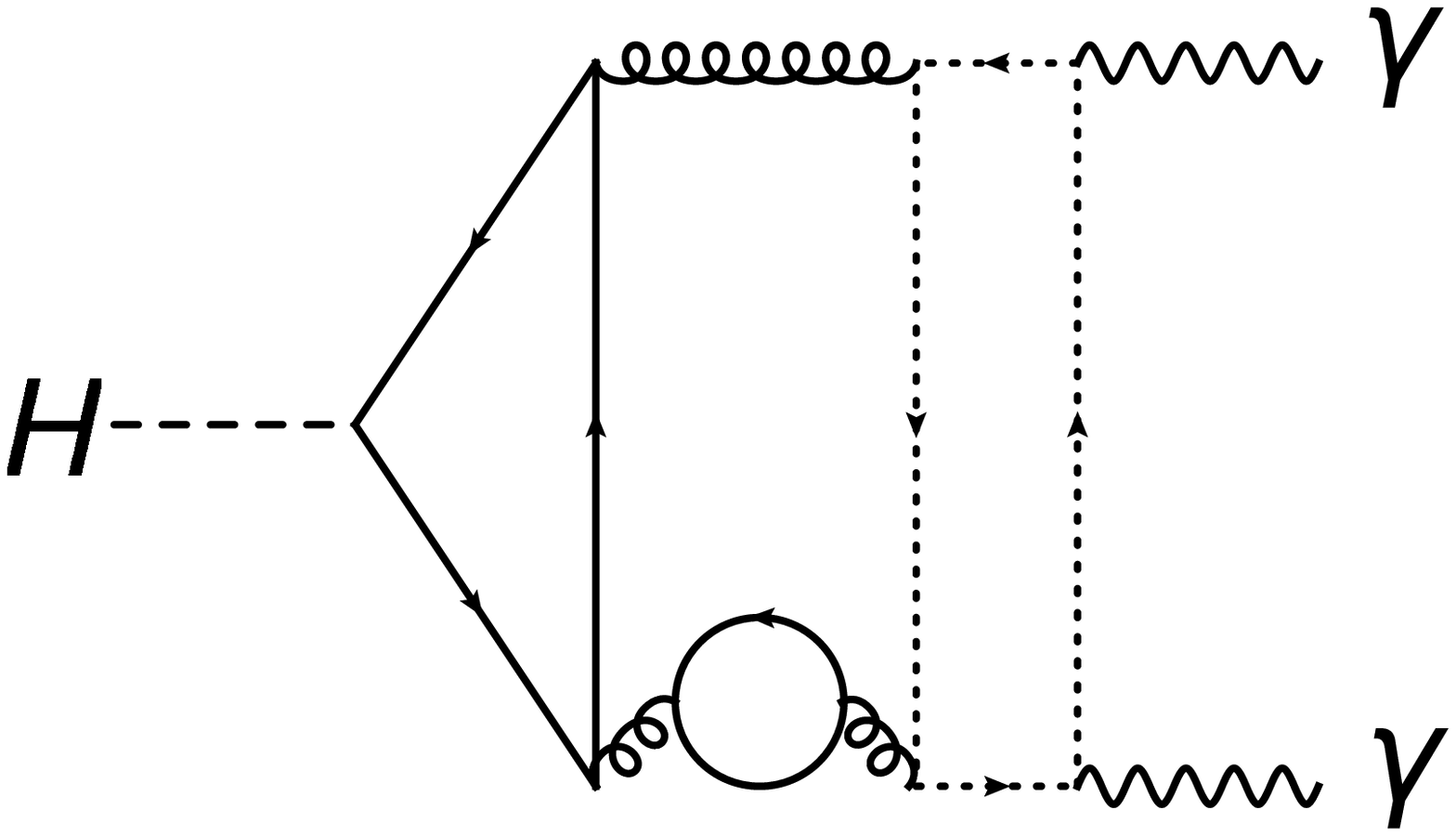}\\
$\si\,\nh\nl\cf\tr^2$
\end{center}
\end{minipage}
\begin{minipage}{3cm}
\begin{center}
\includegraphics[bb=72 433 540 720,width=3cm]{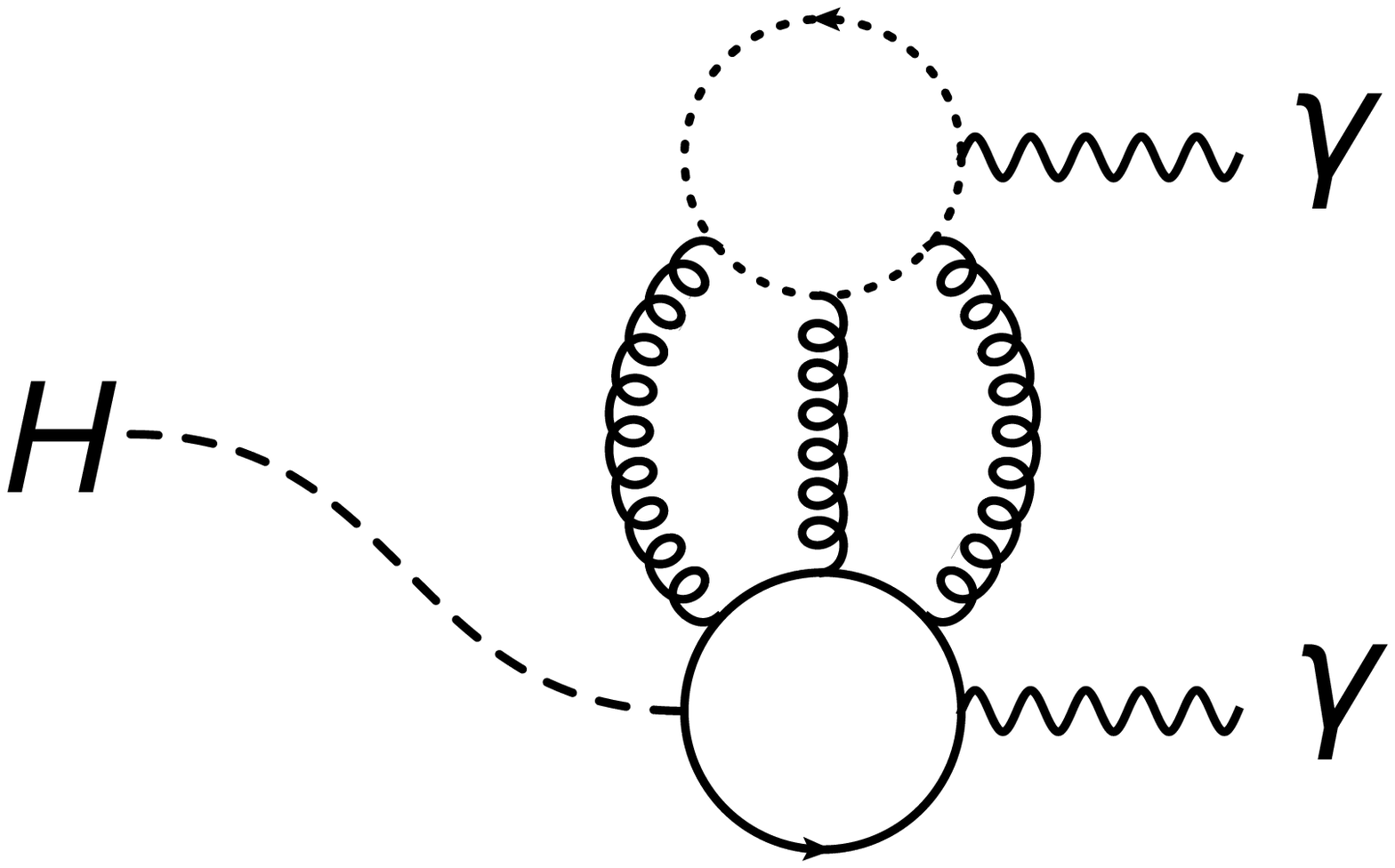}\\
$\nl\dabc\dabc$
\end{center}
\end{minipage}\\[2ex]
\end{center}
\vspace*{-0.4cm}
\caption{Example diagrams where at least one photon couples to a massless
  fermion. Solid lines denote again a heavy top quark, dotted lines
  represent massless quarks, wavy lines are photons, twisted lines
  represent gluons and the dashed line is the Higgs boson.  The
  structure of the color and the flavor insertions of each diagram class
  is given again below each example Feynman graph. 
\label{fig:Simassless4loop}}
\end{figure}

For a Higgs-boson mass of $m_H\approx 126$~GeV the bosonic one-loop
amplitude $A_W$ is real and the QCD corrections to the fermionic
amplitude $A_t$ develop an imaginary part starting from three-loop
order. These imaginary parts have been determined at three-loop order in
Ref.~\cite{Maierhofer:2012vv}.  At four-loop order the contributions to
the amplitude $A^{(3)}$ which originate from the diagrams shown in
Fig.~\ref{fig:Simassless4loop} and those labeled with $\si$ in the last
line of Fig.~\ref{fig:4loop} can in general develop an imaginary part
due to a massless cut. We have checked by using the {\tt{MATAD}} and 
{\tt{MINCER}}~\cite{Steinhauser:2000ry,Gorishnii:1989gt,Larin:1991fz}
routines that they do not appear in Eq.~(\ref{eq:A3}) for the diagram
classes in the last line of Fig.~\ref{fig:4loop}. One can expect that
they are also suppressed by higher powers of the mass correction
$m_H^2/(4\overline{m}_t^2)$ like at three-loop order.

We have also used the logarithmic parts of Eqs.~(\ref{eq:C3qt}) and
(\ref{eq:C3qq}) in order to derive the contributions which are
proportional to the charge factor $Q_{q_i}$ of the light quarks and find
agreement with the result of Ref.~\cite{Chetyrkin:1997un}. The
determination of the real and imaginary part of the diagrams of
Fig.~\ref{fig:Simassless4loop} is beyond the scope of this work.  In
particular the imaginary part of the QCD four-loop amplitude contributes
to the partial decay width only beginning at order $\alpha_s^5$, so that
we do not consider them here.

Finally at five-loop order we determine first the scale dependent part
of the photon vacuum polarization function in the limit of a heavy
top-quark mass with the help of the RGE of Eq.~(\ref{eq:anodim}). From
this result we derive the five-loop contribution
$A^{(4)}|_{\mbox{\tiny{$\gamma tt$-approx}}}$ of Eq.~(\ref{eq:Atexp}) to
the amplitude of the Higgs-boson decay into two photons which arises
from those diagrams where both photons couple to a massive top-quark
loop.  The result is quite lengthy, so that we present it here
explicitly for the $SU(3)$ color group and with the labels
$\nh=\si=1$. It reads
\begin{eqnarray}
\label{eq:A4}
\left.A^{(4)}\right|_{\mbox{\tiny{$\gamma tt$-approx}}}\!\!\!\!\!\!\!\!\!\!\!\!\!\!\!\!\!&=&
- {3460281373\over16329600} 
- {5084\over27}\*\A5 
- {1597771\over3240}\*\A4 
+ {3313\over144}\*\z5
+ {1163954353\over2419200}\*\z3 
- {3699137\over9331200}\*\pi^4 
\nonumber\\&&
+ {1271\over810}\*\logtwo^5 
- {1597771\over77760}\*\logtwo^4 
- {1271\over486}\*\logtwo^3\*\pi^2 
+ {1597771\over77760}\*\logtwo^2\*\pi^2 
- {39959\over19440}\*\logtwo\*\pi^4 
- {29791\over1728}\*\LmusdmMSbars^3 
\nonumber\\&&
- {17701\over864}\*\LmusdmMSbars^2 
- \LmusdmMSbars\*\left({362899\over1296} - {237925\over1152}\*\z3\right)
- \nl^3\*\left[
  {487\over5832} 
- {\z3\over18}
- {\LmusdmMSbars^3\over216} 
+ {\LmusdmMSbars^2\over216} 
- {19\over648}\*\LmusdmMSbars 
       \right] 
\nonumber\\&&
+ \nl^2\*\left[
  {31643\over 20736} 
+ {17\over 54}\*\A4
+ {3319\over 3456}\*\z3
- {4097\over 155520}\*\pi^4 
+ {17\over 1296}\*\logtwo^4 
- {17\over 1296}\*\logtwo^2\*\pi^2 
\right.\nonumber\\&&\left.
- {31\over 144}\*\LmusdmMSbars^3 
- {35\over 288}\*\LmusdmMSbars^2
- \LmusdmMSbars\*\left({7817\over 3456} - {125\over 288}\*\z3\right)
      \right]
+ \nl\*\left[
  {6837097\over 403200} 
+ {328\over 27}\*\A5
+ {10909\over 405}\*\A4
\right.\nonumber\\&&\left.
- {1619\over 216}\*\z5
- {35509969\over 604800}\*\z3 
+ {126527\over 291600}\*\pi^4 
- {41\over 405}\*\logtwo^5 
+ {10909\over 9720}\*\logtwo^4 
+ {41\over243}\*\pi^2\*\logtwo^3 
\right.\nonumber\\&&\left.
- {10909\over 9720}\*\logtwo^2\*\pi^2 
+ {1289\over 9720}\*\logtwo\*\pi^4
+ {961\over 288}\*\LmusdmMSbars^3 
+ {829\over 192}\*\LmusdmMSbars^2 
+ \LmusdmMSbars\*\left({166877\over 3456} - {1925\over 96}\*\z3\right)
     \right]\,.
\end{eqnarray}

\begin{figure}[ht!]
\begin{center}
\begin{minipage}{12cm}
\includegraphics[width=12cm,bb=0 0 567 384]{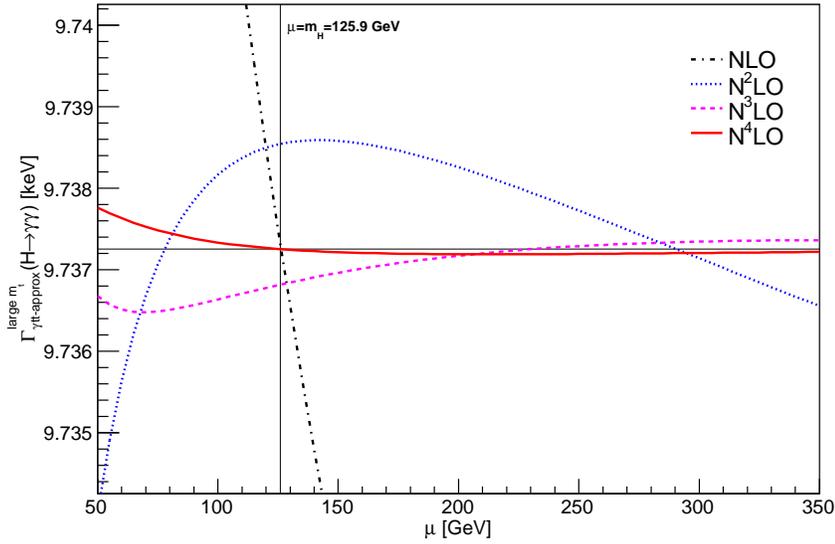}
\end{minipage}
\end{center}
\caption{The dash-dotted,
dotted, dashed and solid lines represent the inclusion of the two-
(NLO), three- (N$^2$LO)
four- (N$^3$LO) and five-loop (N$^4$LO) contribution in QCD to the
quantity $\Gamma^{\mbox{\tiny{large} }m_t}_{\mbox{\tiny{$\gamma tt$-approx}}}(H\to\gamma\gamma)$. The scale has been varied
between 50 and 350~GeV which covers the complete mass scale range $m_H/2<\mu<2m_t$.
The vertical line shows the location where the scale $\mu$ is equal to
the mass of the Higgs boson $m_H$ and the horizontal line gives the
value of $\Gamma^{\mbox{\tiny{large} }m_t}_{\mbox{\tiny{$\gamma
      tt$-approx}}}(H\to\gamma\gamma)$ for $\mu=m_H$ at N$^4$LO accuracy.
\label{fig:scaledep}}
\end{figure}

In order to study the size of the different contributions we perform a
numerical evaluation of the amplitude and derive the partial decay
width. We start with the QCD corrections to the fermionic amplitude
$\hat{A}_t$.  As input parameters we use for the top-quark mass
$\overline{m}_t(m_H)=166.79$~GeV and for the Higgs-boson mass
$m_H=125.9$~GeV~\cite{Beringer:1900zz}.  We obtain for $\mu=m_H$ and the
$SU(3)$ color group
\begin{eqnarray}
\label{eq:Ainftytmumh}
A^{\infty}_t= \hat{A}_{t}\*\Big[ 
      1 
\!&\!\!-\!\!&\!\asH
      -        \asH^2\*( 0.307 + (0.908 - 1.440\*i)_{\mbox{\tiny{si}}}) 
\nonumber\\
\!&\!\!+\!\!&\!\asH^3\*(6.456+(0.746+\tilde{c}_3)_{\mbox{\tiny{si}}})
       -       \asH^4\*(50.808+\tilde{c}_4)
       + \mathcal{O}\left(\asH^5\right)
                  \Big]\,,
\end{eqnarray}
with $\asH=\alpha_s(\mu=m_H)/\pi$.  The index si indicates that the
contributions come from singlet diagrams.  The symbols $\tilde{c}_3$ and
$\tilde{c}_4$ stand for the yet unknown contributions at four- and
five-loop order which arise from diagrams where at least one external
photon couples to massless fermions.  For the scale
$\mu=\overline{m}_t(\overline{m}_t)$ we obtain similarly
\begin{eqnarray}
\label{eq:Ainftytmumt}
A^{\infty}_t= \hat{A}_{t}\*\Big[  1 
\!&\!\!-\!\!&\!\asT 
       -       \asT^2\*(1.292+(0.889-1.440\*i)_{\mbox{\tiny{si}}} )
\nonumber\\
\!&\!\!+\!\!&\!\asT^3\*(5.937+(0.992+\breve{c}_3)_{\mbox{\tiny{si}}})
       -       \asT^4\*(23.220+\breve{c}_4)
       + \mathcal{O}\left(\asT^5\right)
                  \Big]\,,
\end{eqnarray}
with $\asT=\alpha_s(\mu=\overline{m}_t(\overline{m}_t))/\pi$.  At
three-loop order the complete singlet contributions are known and
sizable~\cite{Maierhofer:2012vv}. Depending on the choice of the
renormalization scale they can become approximately as big as the
non-singlet ones, like in Eq.~(\ref{eq:Ainftytmumt}), or up to a factor
three larger, like in
Eq.~(\ref{eq:Ainftytmumh})~\cite{Maierhofer:2012vv}.  One can expect
that the same holds at four- and five-loop order. In the following we
will use the new contributions in order to study the reduction of the
scale dependence. For that purpose we determine the decay width
$\Gamma^{\mbox{\tiny{large} }m_t}_{\mbox{\tiny{$\gamma
      tt$-approx}}}(H\to\gamma\gamma)$, which contains only the leading
contributions in the heavy top-quark mass limit for the diagrams, where
the photons always couple to top quarks only. The dependence of
$\Gamma^{\mbox{\tiny{large} }m_t}_{\mbox{\tiny{$\gamma
      tt$-approx}}}(H\to\gamma\gamma)$ on the renormalization scale
$\mu$ up to five-loop order is shown in Fig.~\ref{fig:scaledep}.  Going
from one-loop to five-loop order the scale dependence continuously
decreases.

The mass corrections and singlet contributions, which are not contained
in Fig.~\ref{fig:scaledep}, are known up to three-loop order.  In the
next step we include them too.  The mass corrections decrease the decay
width.  For the numerical evaluation we use the following input
parameters for the fine-structure constant
$\alpha=1/137.035999074$~\cite{Beringer:1900zz} and the strong coupling
constant $\alpha_s(M_Z)=0.1185$~\cite{Beringer:1900zz}, which we evolve
to the Higgs-boson mass scale $\mu=m_H$.  For the partial decay width
including the QCD corrections we obtain for
$m_H=125.9$~GeV~\cite{Beringer:1900zz}
\[
\Gamma(H\to\gamma\gamma)=
  9.370_{\mbox{\tiny{1-loop}}} 
+ 0.168_{\mbox{\tiny{2-loop}}} 
+ 0.008_{\mbox{\tiny{3-loop}}}  
- 0.002_{\mbox{\tiny{4-loop,$\gamma tt$-approx}}}
+ 0.0004_{\mbox{\tiny{5-loop,$\gamma tt$-approx}}} 
\mbox{ keV}\,,
\]
where each term corresponds to the next order in the perturbative
expansion.  The four- and five-loop orders contain only the
contributions where the photons couple to a top quark in the heavy
top-quark mass limit which is considered in this work. Their size can
change once the complete result becomes available.
Its calculation is beyond the scope of the present work.\\
We convert the top-quark mass to the on-shell
scheme~\cite{Tarrach:1980up,Gray:1990yh,Chetyrkin:1999ys,Chetyrkin:1999qi,Melnikov:2000zc,Melnikov:2000qh}
and include also the two-loop electroweak corrections of
Refs.~\cite{Actis:2008ts,Passarino:2007fp}. For the Fermi-coupling
constant we use $G_F=1.1663787\cdot10^{-5}\mbox{
  GeV}^{-2}$~\cite{Beringer:1900zz} and obtain
\begin{eqnarray*}
\Gamma(H\to\gamma\gamma)&=&[
  9.384_{\mbox{\tiny{1-loop}}} 
+ ( 
  0.159_{\mbox{\tiny{QCD}}}
- 0.150_{\mbox{\tiny{EW}}}
  )_{\mbox{\tiny{2-loop}}} 
+ 0.004_{\mbox{\tiny{QCD, 3-loop}}}
\nonumber\\&&
- 0.001_{\mbox{\tiny{QCD, 4-loop,$\gamma tt$-approx}}} 
+ 0.0006_{\mbox{\tiny{QCD, 5-loop,$\gamma tt$-approx}}} 
                           ]  
\mbox{ keV}
=9.396\mbox{ keV}\,.
\end{eqnarray*}
The different terms corresponds to the one-loop, two-loop and three-loop
contribution, where we have subdivided the two-loop term again into QCD
and electroweak(EW) corrections.  The partial decay width changes by
about $\sim0.13$~keV when the Higgs-boson mass is varied within its
current uncertainty of $\pm0.4$~GeV.
\section{Summary and conclusion\label{sec:DiscussConclude}}
The decoupling function $\zeta_{g\gamma}^2$ relates the $\MSbar$
renormalized fine-structure constant in the full theory with $\nf=\nl+1$
active quark flavors to the $\MSbar$ renormalized fine-structure
constant in an effective theory with $\nl$ light active quark
flavors. We have computed the complete decoupling function
$\zeta_{g\gamma}^2$ at four-loop order in perturbative QCD for a general
$SU(\nc)$ color group as a new result. The decoupling function enters
into the description of an effective Higgs-photon-photon coupling in the
heavy top-quark mass limit. As an application of the calculation we have
used the effective theory in order to determine the four-loop QCD
corrections in the heavy top-quark mass limit to the decay amplitude
$H\to\gamma\gamma$ for those contributions for which the photons couple
directly to the top quark.  The calculation of the amplitude agrees with
a known, independent result in literature.  In addition we also extended
this calculation to five-loop order by using the anomalous dimensions in
combination with the renormalization group equation of the vacuum
polarization function.  Finally we study the reduction of the scale
dependence and perform a numerical evaluation of the partial decay
width. Its dominant uncertainty arises from the error in the Higgs boson
mass.\\

\vspace{2ex}
\noindent
{\bf Acknowledgments}\\
C.S. would like to thank Johann K{\"u}hn and Matthias Steinhauser
for valuable discussions and comments on the manuscript.\\[2ex]
\noindent
The Feynman diagrams were drawn with the help of
{\tt{Axodraw}}\cite{Vermaseren:1994je} and
{\tt{Jaxodraw}}\cite{Binosi:2003yf}.

\begin{appendix}
\section{The decoupling function up to three-loop order\label{app:Pi0}}
The results for the decoupling function
$\zeta_{ph}^{(k)}(\overline{m}_t,\mu)$ of Eq.~(\ref{eq:zeta0}) up to
three-loop order are known from Ref.~\cite{Chetyrkin:1997un} and read
\begin{eqnarray}
\label{eq:C00}
\zeta_{ph}^{(0)}(\overline{m}_t,\mu)&=&-Q_t^2\*{4\over3}\*\LmusdmMSbars\,,\qquad\qquad
\zeta_{ph}^{(1)}(\overline{m}_t,\mu)=-Q_t^2\*\cf\*\left({13\over12} - \LmusdmMSbars\right)\,,\\
\label{eq:C02}
\zeta_{ph}^{(2)}(\overline{m}_t,\mu)&=&Q_t^2\*\Bigg[
   \cf^2\*\left(- {97\over72} 
                + {95\over48}\*\z3 
                - {9\over8}\*\LmusdmMSbars
         \right) 
 + \ca\*\cf\*\left(  {5021\over2592} 
                   - {223\over96}\*\z3 
                   + {7\over9}\*\LmusdmMSbars 
                   + {11\over24}\*\LmusdmMSbars^2
            \right)
\nonumber\\
&-&\nl\*\cf\*\tr\*\left(  {361\over648} 
                       - {\LmusdmMSbars\over9} 
                       + {\LmusdmMSbars^2\over6}
                 \right)
+ \nh\*\cf\*\tr\*\left(  {103\over324} 
                       - {7\over16}\*\z3 
                       + {\LmusdmMSbars\over9} 
                       - {\LmusdmMSbars^2\over6}
                \right)\Bigg]
\nonumber\\
&-&\sum_{i=1}^{\nl}Q_{q_i}^2\*\nh\*\cf\*\tr\*\left(
                            {295\over648} 
                          - {11\over36}\*\LmusdmMSbars 
                          + {\LmusdmMSbars^2\over6}
                           \right)\,.
\end{eqnarray}
The symbols which appear in Eqs.~(\ref{eq:C00}) and (\ref{eq:C02}) are
defined in Section~\ref{sec:results}.

\section{Fermionic amplitude in the heavy top-quark mass limit
at two- and three-loop order \label{app:A1toA2}}
For completeness we give here the result for the fermionic amplitude at
two- and three-loop order in the heavy top-quark mass limit
\begin{eqnarray}
\label{eq:A1}
A^{(1)}&=&-{3\over4}\*\cf \,,\\
\label{eq:A2}
A^{(2)}&=&
  {27\over32}\*\cf^2 
- \cf\*\left[
  \ca\*\left({7\over12} + {11\over16}\*\LmusdmMSbars\right) 
+ \nh\*\tr\*\left({13\over48} - {\LmusdmMSbars\over4}\right) 
+ \nl\*\tr\*\left({1\over12} - {\LmusdmMSbars\over4}\right)
      \right]
\nonumber\\
&+&\si\,\*\cf\*\tr\*\left[
 \nh\*{3\over16}
+Q_t^{-2}\*\sum_{i=1}^{\nl}Q_{q_i}^2\*
\left(\z3-{13\over8}+{1\over4}\*\Log{-{m_h^2\over\overline{m}_t^2}}\right)\right]
\,,
\end{eqnarray}
where we have taken the last terms, which are proportional to
$Q_{q_i}^2$, from Ref.~\cite{Maierhofer:2012vv}.  The symbols which
arise in Eqs.~(\ref{eq:A1}) and (\ref{eq:A2}) are again defined in
Section~\ref{sec:results}.

\providecommand{\href}[2]{#2}\begingroup\raggedright\endgroup

\end{appendix}
\end{document}